\documentclass{article}
\usepackage{bm}
\usepackage{MnSymbol}
\usepackage{booktabs}
\usepackage[ruled,vlined]{algorithm2e}
\usepackage[british]{babel}
\usepackage{arxiv}
\usepackage{amsmath}
\usepackage{amsfonts}
\usepackage{amsthm}
\usepackage{graphicx}

\makeatletter
\newcommand{\hbbar}{%
  \text{\m@th
    \makebox[0pt][l]{\raisebox{0.05\height}{$\mathchar'26$}}%
    \makebox[0pt][l]{\raisebox{-0.12\height}{$\mathchar'26$}}%
  $h$}
}
\makeatother

\newcommand{\squisim}{\rotatebox[origin=c]{-60}{\reflectbox{$\bm{\squigarrownwse$}}}}

\newcommand{\squisimscal}{\rotatebox[origin=c]{-60}{\reflectbox{$\squigarrownwse$}}}

\newcommand{\sintrc}[3]{\ensuremath{\squisimscal{#1}^{#2, #3}}}

\newcommand{\intrc}[3]{\ensuremath{\squisim_{#1}^{#2, #3}}}

\newcommand{\rintrc}[3]{\ensuremath{\not \squisim_{#1}^{#2, #3}}}

\theoremstyle{definition}
\newtheorem{definition}{Definition}

\title{A Generalised Theory of Interactions - II. The Theoretical Construction}

\author{
Santiago N\'u\~nez-Corrales$^{1*}$ and Eric Jakobsson$^{2\dagger}$ \\
$^{1}$Illinois Informatics and National Center for Supercomputing Applications,\\ University of Illinois at Urbana-Champaign. Urbana IL, USA.\\
$^{2}$Deparment of Physiology, Molecular and Cellular Biology and National Center for Supercomputing Applications, \\University of Illinois at Urbana-Champaign. Urbana IL, USA.\\
* Corresponding author \texttt{nunezco2@illinois.edu} \\
$\dagger$ In memoriam.
}

\begin{document}

\maketitle

\begin{abstract}
After discussing the significance of interactions to understand complex multiscale stochastic systems (CMSS), we turn our attention to the construction of a Generalised Theory of Interactions (GToI). We define interactions as discrete, localised events where the thermodynamically irreversible exchange of degrees of freedom involves signal propagation and relaxation phenomena. We proceed to define the mechanics of such space, as well as how these approximately map onto dynamical manifolds. By characterising various systems through the effect of transformations across interaction space, we develop a view of systems where features such as emergence and information representation become apparent. Our theory appears to capture structural and dynamical features of CMSS efficiently by making explicit the extent of the dynamical range of action in them; we identify the boundaries of such range with statistical limits corresponding to physical laws, and  derive a generalised form of the Correspondence Principle.
\end{abstract}

\section{Requirements towards a GToI}

For the GToI to successfully overcome several of the shortcomings present across scientific theories described in the first manuscript of this series \cite{nunez2020gtoia}, its underlying core principles must provide sufficient tools to avoid them from the start. Our reasoning follows that by Lee Smolin \cite{smolin2013time} at the top level by requiring our theory to (a) accurately capture systems across a span of complexity that is already well known, (b) generate empirically  falsifiable statements, (c) provide means to understand the  "why these laws?" questions for specific instances and problems, and (d) provide testable answers to the "why these initial conditions?" question. These general statements must therefore correspond to more specific accounts of entities, dynamics and laws.

\subsection{Epistemological requirements}

\textbf{Explanatory closure.} CMSS, although involving multiple levels of action and hierarchical organization, should be renormalizable as \emph{an interaction between systems and their environment} and be complete up to a reasonable approximation. When this sort of renormalization preserves empirical observations, its contents should only contain descriptions of those two entities. At various levels of granularity, the effect of renormalization should be reflected on the level of detail of the outcomes. In our view, however, explanatory closure does not imply explanatory exclusion \cite{gibb2009explanatory}: multiple explanations based on a GToI may be complete and causally consistent without resorting to postulating other external events or unverifiable interactions. Such a type of explanatory closure has been extensively discussed in the context of the landscape problem for string theories in cosmology \cite{susskind2003anthropic, smolin2013perspective}.

\textbf{Relationality.} Relational approaches have proven to be fruitful multiple times across the history of physics \cite{huggett2018absolute} and science in general. A theory is relational if the degrees of freedom involved in the description of an entity involved in a given phenomenon depends solely on the degrees of freedom of other entities without requiring any pre-existing, global medium. We seek a theory where both entities and governing laws emerge as a consequence of the occurrence, aggregation and composition of interaction; presupposing the existence of a containing background not only limits the reach of such endeavour but eliminates its generality and potential for causality across all scales. A relational approach may permit more easily to crack open the highly contextual whole-environment coupling that characterises descriptions of emergence \cite{corning2002re}. More generally, general systems theory \cite{boulding1956general,mesarovic1975general}, with an emphasis on where some sort of evolution is possible \cite{ehresmann1987hierarchical}, admit (or rather demand) relational descriptions. Finally, theories stated in relational terms tend to possess useful compositional properties 743..3.0/even in the most intuitive representation (i.e. set theory vs category theory or topos theory \cite{awodey2003relating}) that usually lead to the discovery of non-trivial properties of phenomena --such as invariants- that facilitate reasoning about them, since these tend to arise constructively as intermediate steps or byproducts towards abstract descriptions of systems and their dynamics --e.g. \cite{isham1997topos}.

\textbf{Background independence.} We naturally expect the GToI to apply to theories where fields and particles of various sorts are involved, and in particular to the form of any experimental theory quantum theory of gravity if such arises. A strong case has emerged in favour of background-independence \cite{smolin2006case}, which has been called ``the most momentous requirement a quantum theory of gravity must satisfy'' \cite{becker2014route} since it mandates an emergent view of space-time that compels certain unifications to occur. Background independence should not be an impediment for the existence of renormalization entities  \cite{reuter2002renormalization,ohta2017background} that may provide the basis for a background-dependent description with asymptotic safety --i.e. avoidance of infinities when computing well-defined extensive quantities \cite{niedermaier2006asymptotic}. Thus, when the difference between interaction scales is sufficiently large for two classes of events, at least one background-dependent description must exist. We note that, while this description is centered on theories of quantum gravity, its application can be extended to other areas, including social science \cite{ike2016background}.

\textbf{Locality.} Across most physical theories, locality characterises the separability and independence of events across space-time due to the finite velocity at which information can travel \cite{bitsakis1990locality}: locality acts as an \emph{selector} operator that restricts the number and value ranges of degrees of freedom where observables acquire values, to sets of finite measure, or in the case of point particles or processes, sets of zero measure. This manifests itself clearly across field theory: the selection process corresponds to replacement of the algebra of observables when the region being measured changes \cite{roberts1982localization}. More specifically, it is a statement about the upper limit of information obtainable by means of measurements, as well as of the measurements' perturbation effect on systems. Locality is crucial in the interpretation of the relevance of acceleration when measurements are performed within a relativistic frame of reference \cite{mashhoon1990hypothesis}. If locality is more deeply injected into general relativity by curving the geometry of the energy-momentum space, non-commutativity of momentum and non-linear conservation laws arise; both are encouraging (and relationally consistent) with expectations for a theory of quantum gravity \cite{amelino2011principle}.  Stating the GToI in terms of locality should help reconcile opposing views between theories that postulate the existence of global, cosmological laws and those that see the latter as practically convenient but epistemologically unsound extrapolations \cite{smeenk2017cosmos}. By calculating the consequences of assuming only locality and probabilistic measurement operators as the basis for physical action, quantum-like laws emerge \cite{oeckl2014first}. Similarly, it has been shown that the presence of unitary operators in the algebras describing Lorentz-invariant time evolution while avoiding certain undesirable outcomes (e.g. vacuum instability, ultraviolet catastrophe) is only possible when the corresponding Hamiltonian is stated in terms of local interactions \cite{guenin1966interaction}. Finally, locality is used to define, by contraposition, non-separable events in space-time, or non-locality.

\textbf{Causality.} Spatial relativity places a universal limit on signal propagation that holds in general relativity \cite{carter1971causal} in order to preserve Lorentz invariance: that is, causality is embedded in the structure of space-time. Causality can take the form of constraints of the solutions of dynamical systems \cite{souza2002dynamics}. The situation appears no different either in quantum mechanics when referring to measurements performed on systems with interaction observables \cite{dimock1980algebras} such as scattering cross sections \cite{papp1972interaction,cufaro1981action}, in situations that involve non-locality such as in EPR pairs \cite{popescu1998causality}, or in quantum field theories for local propagation phenomena (i.e. conservation of microcausality \cite{alebastrov1974causality}), and causality violations are sought after in high-energy physics as the signature of composite structures for particles \cite{joglekar2008composite}. If causality is interpreted as a non-empty set of conditions resulting from the interaction of two physical processes \cite{salmon1984scientific,ehring1986causal}, a mechanistic explanation is required to understand how that set is selected from other alternatives. While equating causality with interactions is not new \cite{heathcote1989theory}, our view is that the association is in fact universal and applies across scales of action. It has been observed in some hard cases --e.g. the Fokker-Wheeler-Feynman retarded field action \cite{moore1992causality}- causality can be both preserved formally and made sense physically. When we move up the complexity latter, causality becomes harder to establish \cite{ellis2005physics,ellis2008nature}. In biology, robustness introduces degeneracy and redundancy in ways that make difficult the establishment of causal relations \cite{lesne2008robustness}, a situation which is only made more clear across systems of entities with adaptation and learning capabilities \cite{sliva2019combining}. Causality in conjunction with entropy defines a natural separation between classes of stochastic processes \cite{zunino2010complexity}. Methodologically, the analytic deconstruction and reconstruction of complex phenomena by means of models should not change causality relations \cite{hogan1987modularity}, but it should be possible to unveil the origin of causal relations when these have indirect origins, e.g. dynamical couplings \cite{sugihara2012detecting}, and, consequently, causality in a GToI should acquire a strongly topological flavour \cite{harnack2017topological}. In our approach, only pairs of interacting entities are allowed.  Many-body interactions emerge from the underlying pairwise interactions.

\textbf{Thermodynamic irreversibility.} Departing from systems in steady thermal states is often motivated by the difficulties entailed in the treatment of systems far from equilibrium where irreversible action occurs. A successful GToI should not only integrate thermodynamical irreversibility, but be directly built upon it as a more natural --and hopefully more correct- description of systems with interactions \cite{prigogine1978time}. The resolution of several paradoxes across various disciplines and their rooting in a common framework that ties them to gravitation would constitute positive evidence in that direction \cite{gal1976cosmological}. We note that, interestingly, describing thermodynamics by means of Riemannian geometry leads to the appearance of non-zero curvature in the respective manifold when interactions are introduced \cite{ruppeiner1979thermodynamics}, echoing the appearance of curvature in general relativity when massive objects are introduced in space-time.  Interactions between two non-equilibrium systems with a shared interface has been studied in composite entities \cite{muschik2004thermodynamic}, but no immediate path toward generalisation becomes evident. Endoreversible thermodynamics provides a better model for our purposes, since irreversibility is explained in terms of interactions with asymmetric heat exchange between systems whose components may be reversible \cite{wagner2015chemical}. The extension of irreversible thermodynamics to networks \cite{oster1971network} brings us, topologically, a step closer to what we expect in connection to causality. A fully integrated view of interactions and thermodynamic irreversibility requires a generalised form of uncertainty: the existence of uncertainty relations in classical mechanics analogous those found in quantum mechanics \cite{furth1933einige} motivates this assertion. One challenge a GToI should help further clarify of is causal relationship between energy and temperature in statistical mechanics \cite{kalinin2005boltzmann}. Another challenge corresponds to providing better reasoning around the frequent observation of core-halo distribution formation in systems far from equilibrium beyond symmetry breaking of ergodicity \cite{levin2014nonequilibrium}; finding a compelling explanation under a more general framework may lead to significant generalisations applicable to aggregate and composite systems. From the perspective of symmetry breaking \cite{prigogine1973breaking}, irreversibility appears to be strongly associated to the ratio between microscopic and macroscopic scales \cite{lebowitz1999statistical}; simultaneously, outcomes obtained using a GToI should not focus on specific initial conditions, but rather on ensembles of initial conditions and probably their partitions. 

Using both statistical thermodynamics and microphysics approaches, one can find situations where aggregates move up thermal gradients in apparent contradiction of laws known for simpler systems \cite{dhont2004thermodiffusioni,dhont2004thermodiffusionii}: thermodynamic irreversibility across CMSS brings puzzling questions that, once more, a GToI should better help elucidate. The role of irreversibility in biology was identified early \cite{prigogine1946biologie}. For instance, organismal development correlates with a decrease in the rate of entropy production \cite{zotin1967thermodynamic}. Stochasticity that arises from it in biological systems endows them with flexibility and adaptation, both responsible for functional robustness \cite{lesne2008robustness}.  In production theory in economics, time irreversibility does not necessarily involve thermodynamic irreversibility, even if both are well defined for a particular case \cite{baumgartner2005temporal}; this is significant since the arrow of time --i.e the irreversible character of time as a physical dimension- is built using thermodynamic irreversibility arguments. More generally, a GToI should anticipate the analysis of fractal systems that arise from dynamics acting on dense non-continuous sets \cite{sokolov2001thermodynamics}; the existence of methods to find order parameters and the phase transitions these encode is suggestive of the existence of similar possibilities when reasoning in terms of interactions \cite{bohr1987order}. We suspect that rigorously describing \emph{reversible} systems using a GToI will yield concrete theories of interaction that are either cumbersome (proportional to the variety of interactions involved) or conspicuously unphysical unless additional assumptions are added. This situation would mirror some of the difficulties when describing irreversible systems in current approaches.

\textbf{Compositionality.} Compositionality appears to be the simplest mode of organization that attains complexity --i.e. an outcome different from pure aggregation- when departing from a substrate of interacting entities \cite{wimsatt1994ontology}. In addition, as an epistemological device, it can provide convenient abstractions provided the underlying ontology partitions the space of structures and events into tractable units \cite{strevens2017ontology}. Compositionality tends to result in elegant algebraic treatment of difficult subjects. A GToI therefore must carefully choose its entities and compositional laws as to remain empirically testable and intellectually economic with an algebraic bent: for many systems, intuitions will be hard to come by and similar to quantum mechanics, and the robustness of the formal expressions and the ability to find invariants will be some of the few assurances available. Compositonality requires in complex systems that emergence occurs by virtue of the existence of composition operators acting both on entities and on the rules governing their dynamics that can be systematically applied \cite{chen2007calculus}; as a result, generative effects must be manifest \cite{adam2017systems}.

Amidst the natural sciences, the significance of compositionality becomes most apparent in biology. Biochemical interactions, including those between enzymes and substrates, can be captured by various algebraic means relying on composition \cite{wright2018bond}. Systems biology has not been the exception \cite{prandi2004shape}. Hierarchical evolutive systems \cite{ehresmann1987hierarchical} and more recently memory evolutive systems \cite{ehresmann2007memory} depend on the compositional language of category theory to express static and dynamical properties of biological systems that change under selective pressures across multiple scales. \textit{Designed} systems constitute an interesting case. Compositionality appears to be a key ingredient for the fluidity of intelligence in cognition \cite{duncan2017complexity}. Computation as a formal process can be described by means of interactions \cite{goldin2006interactive}, for instance, having undergone a complexity explosion only possible due to the compositional properties of hardware and software. In particular, complex distributed software systems can be specified through composable interactions \cite{chilton2013algebraic}. Succinctly, a GToI must provide the scaffolding required, in principle, to reconstruct the large-scale structure of the universe from the most fundamental level possible \cite{smolin1995cosmology}.

\textbf{Computability.} For a GToI computability involves both (a) the ability to obtain numerical outcomes that predict or reconstruct specific situations and (b) the existence of an approximate isomorphism between physical systems and an algebra whose establishment takes a finite number of transformations to realise \cite{feynman2018feynman}. Computing consists of thermodynamic actions, many of which are irreversible \cite{landauer1961irreversibility}; computations along the spectrum of irreversibility suggest an arrow of time emerges when information is erased \cite{adriaans2011computation}. Since computation is a physical act, we expect a GToI to provide adequate estimates for the limit of what a computer can do \cite{lloyd2000ultimate}. Using the rise of quantum computing as an example, understanding existing computers using physics tends to shed new light into what processes may also count as computing besides Turing machines \cite{deutsch1985quantum}; it is likely for a computational theory of interactions to result in other definitions of what it means to compute. To this end, we anticipate that stating computations is a GToI, given its stochastic roots, will likely be performed by some type of bisimulation \cite{larsen1991bisimulation}: a compositional, physical system used to sample spatially, temporally and informationally another system through repeated interactions. We also note that when external interactions are introduced into algorithm analysis, then the computation becomes open --i.e. stops being complete or self-consistent \cite{wegner1997interaction}. If an interactive computation yields useful work, it is therefore out of equilibrium, which implies some arrow of time imposed by thermal or information loss. Since information appears to play a critical critical across fundamental physics \cite{wheeler1992recent}, a GToI should help clarify the processes that produce it in terms of information producing processes and the robotics required to move bits around during computations, and capture causality in topological fashion --e.g. \cite{jalili2017information}.

\subsection{Pragmatic requirements}

What should, in practice, a GToI provide? First and foremost, we seek to establish the foundations of a theory capable of facilitating the description, analysis and testing of hypotheses across CMSS, but also across systems usually not thought of as CMSS. To be consistent with the empirical requirements, a GToI must remain somewhat abstract and parameterizable. Thus, concrete phenomena and/or hypotheses should require \emph{concrete theories of interactions} (CToI). In this sense, the trajectory of the search for a GToI parallels existing work on constructor theory by Deutsch and Marletto, which departs from an abstract specification \cite{deutsch2013constructor} and materialises into probability theory \cite{marletto2016constructor}, information \cite{deutsch2013constructor}, life \cite{marletto2015constructor} and thermodynamics \cite{marletto2016constructor}.

On the side of complexity, the GToI should competently describe interacting systems with ample microstate variety and various degrees of rigidity. We also seek to capture, most importantly, systems with a high degree of \emph{repertoire variety}. Microstate variety describes the number of classes to which the constituent interacting entities belong at the most immediate microscale. For instance, a classical interacting gas with one species has low microstate variety, while interacting proteins in the cell environment constitute a system with high microstate variety. Rigidity \cite{goldbart2002rigidity} entails (a) the stability of macroscale identity under microscopic perturbations --i.e. entity substitution, changes in local topology-, (b) conversion of point-like forces that cause local deformations in microscopic degrees of freedom into global transformations with various continuous macroscopic symmetries, and (c) stability of microscale identity under energy fluxes and entropy production. Repertoire variety corresponds to the number of different types or \emph{modes} of interactions per composing entity in a system, which has significance for a number of reasons. We are interested, for example, in understanding why in some systems emergence only occurs with many simple entities (e.g. conductivity in a copper wire) while in others a few composing entities give rise to rich emergent behaviour. This fact motivates us to hypothesise that the thermodynamic limit need for emergence to manifest is inversely proportional to the repertoire variety of the system.

Multiscale structure should be captured by the GToI, with an emphasis on those observed across CMSS. First, the GToI should describe simple aggregation processes with varying degrees of flexibility. Second, it should help understand the rise of hierarchies and modularity in the presence of dynamical constraints as efficient topological embeddings \cite{bassett2010efficient}. Third, it should provide a model for how noise percolates (and sometimes amplifies) across scales; noise percolation is extremely significant for both simple \cite{visser2011conservative} and complex \cite{wissner2013causal} situations. Finally, a GToI should help broadly capture cross-scale excitability, which is of particular interest both for biology and large-scale distributed information systems. Metabolic networks exemplify all four of these requirements \cite{guimera2005functional,levine2007stochastic}.

Stochasticity in CMSS should appear in various forms in a GToI. It its most general form, it should manifest as uncertainty constraints across all scales \cite{furth1933einige}. As a consequence, the presence of noise should be traceable to either intrinsic sources (e.g. the probabilistic nature of quantum mechanics) or combinatorial causes (e.g. the difficulty of finding the most significant epistatic interactions of an arbitrary number of SNPs that explain the onset and progression of a disease). A GToI should provide the scaffold to quickly model dissipation and fluctuations observed across irreversible processes. In our view, stochasticity is not a useful addition to be preserved only when convenient and otherwise frequently expelled, but a ubiquitous property of the universe as a whole.

To summarise, a GToI is pragmatically successful if it can be used to construct concrete theories of interaction without violating any of the epistemological requirements across systems that range from point-like to fully interconnected, from single-scale to hierarchically modular, and from reversible and closed to irreversible and open. In the following description of an example of such theory, bold symbols indicate non-scalar quantities, and bold capital symbols denote transformations or observables.

\section{Elements of the GToI}

\begin{figure}[htp]
    \centering
    \includegraphics[width=3.5in]{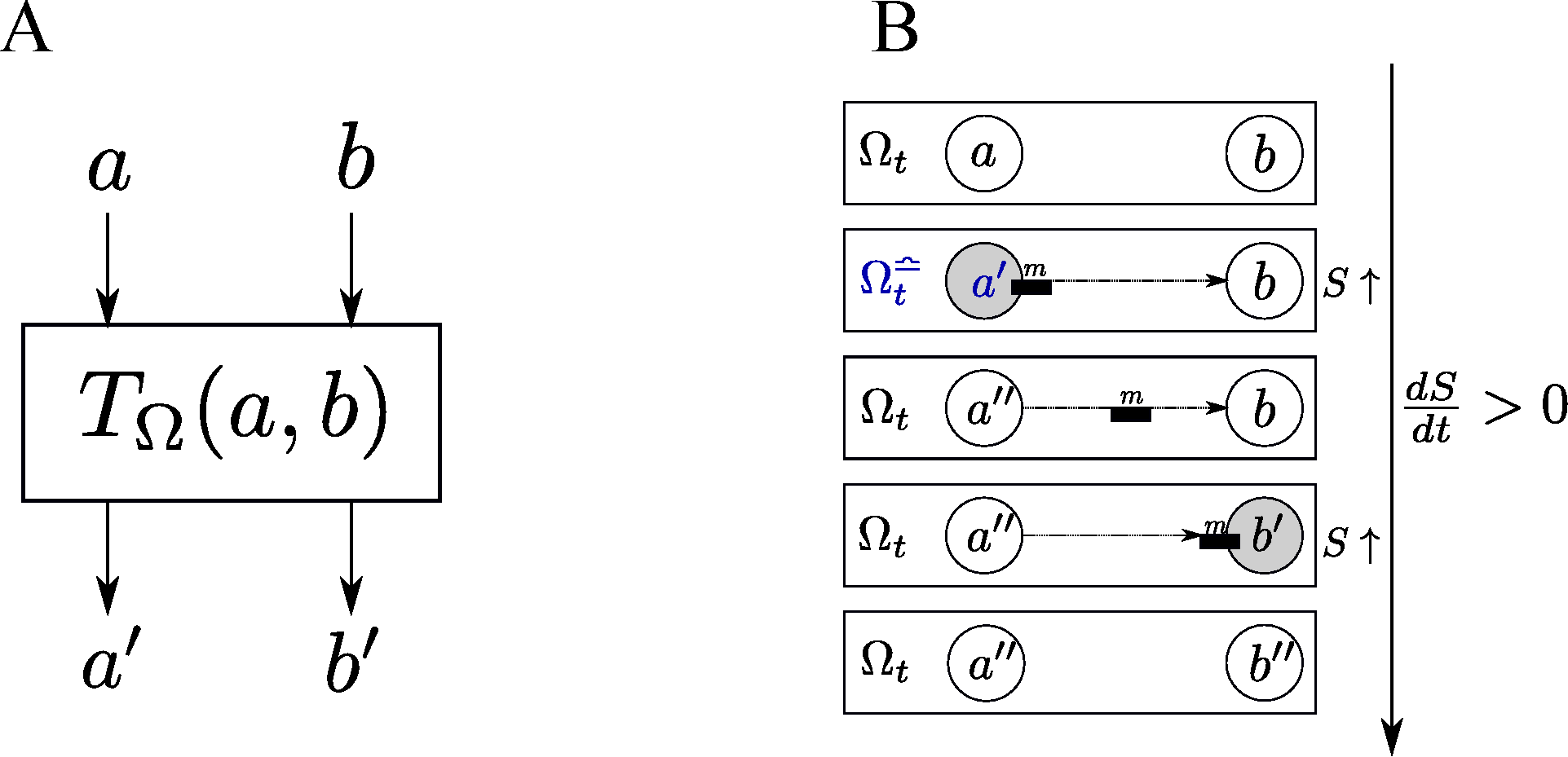}
    \caption{Two abstract models of an interaction. \textbf{A.} The standard transformation model where interactions are inferred from nonlinear changes in properties of $a'$ and $b'$. \textbf{B.} An explicit model of interactions where degrees of freedom are exchanged via a messenger $m$ with entropy production after state relaxation.}
    \label{fig:fig_1}
\end{figure}

Most models of interactions across the sciences resemble in some form or another the situation described by Figure \ref{fig:fig_1}A. For two systems $a,b$ embedded in a manifold $\Omega$ --e.g. spacetime, an electromagnetic field, a heat bath, the interior of a cell, an economy, society- a transformation $T_\Omega$ is defined such that

\begin{equation}
    \bm{T}_\Omega(a, b) = (a', b')
\end{equation}

such that the difference between some quantity $\bm{\Gamma}(a, b, \bm{T}_\Omega)$ yields a nonlinear curve along some dimensions, usually time. Cross sections constitute a paradigmatic example. To compute its consequences, however, the quantity dependent on time and space he locus of interaction is given by the centre of mass

\begin{equation}
    \bm{r}^\mu_{a,b} = \frac{m_a \cdot r_a^\mu + m_b \cdot r_b^\mu}{m_a + m_b}
\end{equation}

where $\mu$ is the coordinate index and $m_a, m_b$ are the respective masses. Furthermore, assume that to every function $\bm{\Gamma}$ stated in terms of entities corresponds a function $\bm{\gamma}$ whose value only depends on $\bm{r}_{a,b}$. For a single interaction in $\Omega$

\begin{equation}
    \bm{\gamma}(\bm{r}_{a,b}) = \int_{\Omega} \delta(\bm{r}_{a,b} - \bm{r}) \bm{\gamma}(\bm{r}) \bm{dr}
\end{equation}

When applied to a position $\bm{r}$ different than the centre of mass, we expect $\bm{\gamma}(\bm{r}) = 0$. Integrating on both sides

\begin{equation}
    \int_\Omega \bm{\gamma}(\bm{r}) \bm{dr} = \iint_{\Omega} \delta(\bm{r'} - \bm{r}) \bm{\gamma}(\bm{r}) \bm{dr} \bm{dr'}
\end{equation}

we obtain a form that, when interactions are sufficiently dense in $\Omega$, suggest the existence of a continuous kernel $\bm{K}(\bm{r'}, \bm{r})$ such that

\begin{equation}
    \iint_{\Omega} \delta(\bm{r'} - \bm{r}) \bm{\gamma}(\bm{r}) \bm{dr} \bm{dr'} \approx \iint_{\Omega} \bm{K}(\bm{r'},\bm{r}) \bm{dr} \bm{dr'}
\end{equation}

which, for sufficiently small $\delta r = ||\bm{r'} - \bm{r}||$, a corresponding kernel $\bm{G}(\bm{r})$ exists with the property

\begin{equation}
    \iint_{\Omega} \bm{K}(\bm{r'},\bm{r}) \bm{dr} \bm{dr'} \approx \int_{\Omega} \bm{G}(\bm{r}) \delta r \bm{dr} \approx \int_\Omega \bm{\gamma}(\bm{r}) \bm{dr}.
\end{equation}

We find in the resulting expression a \emph{position dependent field} $\bm{\Phi}(\bm{r})$. Let us analyse what information we have gained. First, fields can be reconstructed from sufficiently dense collections of interactions, or rather, \emph{interaction effects}. Second, that the field is a variational quantity dependent on $\delta r$. Third, that for the field to be properly defined, suitable functions $\bm{K}$ and $\bm{G}$ must exists, and be continuous and deterministic. Since we are interested in systems with stochasticity, one might be tempted to introduce an integration measure from a random process $\bm{W}_{\bm{r}}$ such that

\begin{equation}
    \bm{\Phi}(\bm{r}) = \mathbb{E} \left( \int_{\Omega} \bm{G}(\bm{r}) \delta r \bm{dW}_{\bm{r}} \right).
\end{equation}

A challenge arises in how to interpret the noise term, which depends on the assumptions about  the generating random process, which may be resolved to some degree of satisfaction by gaining additional information about the system and selecting the scale of events \cite{smythe1983ito,oksendal2013stochastic}. The second challenge is that the introduction of $\bm{W}_{\bm{r}}$ strongly limits the classes of functions $\bm{G}$ for which an analytic result is possible; numerical calculations become no easier in the selection of the solution approximation method or the computational cost. Assuming these obstacles can be overcome, the process has left us with insights about the analytic and numerical properties of the continuous (or stochastic) response of $\bm{\gamma}$ under change of position in the manifold.

What becomes most significant is what we have not learned. While we may get a clear view of interaction effects, we remain in the dark about \emph{what an interaction is} and \emph{what exactly happens during one}. Not only have we abstracted away relevant relational details provided by the position of $a$ and $b$ in the manifold, but hidden all aspects of the processes behind $\bm{T}_\Omega$. The structure of $a$ and $b$ remains opaque under the guise of behaving as point particles. What is their inner structure? Does it change? What is the nature of that change if they do? Is it dependent on the types of the entities involved? Why is it considered effectively instantaneous? What does the thermodynamic landscape underlying $\bm{T}_\Omega$ look like? Can we understand why nonlinearities appear (if any)?

Let us take a different approach and forget altogether --yet only temporarily- about the response $\bm{\Gamma}$, described by Fig. \ref{fig:fig_1}B. We start by endowing $a$ and $b$ with inner structure whose rigidity varies. For the moment, suppose that an entity $a$ can be represented as a weighted connected graph of degrees of freedom $G(a) = G(V, E, W)$. We use graphs as a reasonable model since their rigidity is well-defined \cite{laurent1997cuts,alfakih2000graph} and reason in terms of its edges. In this physical situation, entities $a, b$ exist within a local dynamic context $\Omega_t$, which may contain fluctuations across its degrees of freedom. We define the dimension $D$ of $\Omega_t$ as the number of degrees of freedom that provide a holonomic basis for $G(a), G(b)$ and their embedding into $\Omega_t$. To exert some effect on either $G(a)$ or $G(b)$, we expect $\Omega_t$ to behave locally as a graph whose edges connect vertices of any $G$ with vertices in it on its surface. Since the local context is time dependent, then it must be the case that $G(a)$ and $G(b)$ also are as a function of their rigidity. If stochastic fluctuations occur in $\Omega_t$, these must also filter into the graph of embedded entities. As long as the volume an entity the manifold $\Omega_t$ does not exceed a certain value, we consider stochastic shapes to be well-defined \cite{kendall1989survey}. Assuming a regular shape without loss of generality, we expect its volume and surface relations in terms of edges $E_a$ to follow respectively

\begin{equation}
    V_a \equiv V[G(a)] \propto |E(a)|, \qquad S_a \equiv S[G(a)] \propto |E(a)|^{\frac{D - 1}{D}}.
\end{equation}

Since $G(a)$ embedded within $\Omega_t$ corresponds to a discrete shape, the equivalent of the Stokes theorem in discrete exterior calculus exists \cite{hirani2003discrete}. Equipped with these notions, we aim to describe an interaction in terms of the consequences of perturbations of varying magnitude to the edges --i.e. relational degrees of freedom- that comprise each interacting entity. An interaction between the two entities, when these are at sufficient proximity from one another, is defined as the process where an active entity $a$ receives a perturbation, relaxes its state and yields either a perturbation into $\Omega_t$ or generates a messenger $m$ using its own structure that acts as a perturbation for another (possibly active) entity $b$. Let us suppose that only a small amount of degrees of freedom in $a$ are impacted by a perturbation. The task before us requires providing a model for (a) the magnitude of the response as a function of the magnitude of the perturbation, (b) the propagation of the response across entities and (c) its consequences for an interaction with another entity $b$.

First, we will separate --with some notational abuse- the edges of an entity into three subsets corresponding to a rigid component, a perturbable component and a heat component,

\begin{equation}
    E(a) = E_R(a) + E_\Delta(a) + E_Q(a).
\end{equation}

To address (a), we observe that the effect of $E_\Delta(a)$ is to capture the internal reconfiguration of recognisable degrees of freedom as a means to decrease internal entropy, while $E_Q(a)$ captures the loss of degrees of freedom to the context as fluctuations (i.e. heat radiation). We also expect $E_\Delta(a)$ and $E_Q(a)$ to increase when a perturbation arrives. We expect the magnitude $M_\phi$ of any perturbation $\phi$ to determine the internal distribution of $E(a)$ in terms of its three components as expected by conservation of mass. As a simplistic model, we considered a sigmoid perturbed fraction of edges $f_\phi$

\begin{equation}
    f_\phi(M_\phi) = C_\alpha + \frac{1}{1 + e^{-\alpha M_\phi}}
\end{equation}

with $C_\alpha$ such that $f_\phi(M_\phi/2) = \frac{1}{2}$, thus 

\begin{equation}
    E_\Delta(a) + E_Q(a) = f_\phi(M_\phi) E(a).
\end{equation}

We expect, following intuitions about radiating processes, the perturbable component to be proportional to volume and the heat component to the surface proportional to the perturbed fraction, taking care in removing the latter from the former one

\begin{equation}
    f_\phi(M_\phi) E(a) = V^\Delta_a + S^Q_a
\end{equation}

with $\beta \leq 1$ such that

\begin{equation}
    \label{eq:disp_pot}
    \beta = \frac{V^\Delta_a - S^Q_a}{V^\Delta_a}
\end{equation}

revealing that $E_\Delta(a) = V^\Delta_a - S^Q_a$ and $E_Q(a) = S^Q_a$. We identify $\alpha$ with the reconfigurability of the system, since it provides information about the potential for an internal reconfiguration, and $1 - \alpha$ with its rigidity. The parameter $\beta$ corresponds to the dissipative potential, or the ability to radiate heat by loosing degrees of freedom to the stochastic context. Both quantities depend on the structure and composition of $a$.

Turning our attention to (b), we model the propagation of the action as collections of edge reconfigurations on $E(a)$. Again, without loss of generality, let us suppose the existence of a path $\pi$ of length $\ell$ such that the number of edges involved in the reconfiguration follows a binomial distribution with probability $p$ and $N = \ell$. By choosing to express change in terms of $\ell$ instead of time, the description remains scale-independent. We quickly observe that the choice of $p = \beta$ matches our dynamical intuitions for passive and active systems: passive systems tend to minimize internal entropy by radiating heat, while active systems do so by reconfiguring its internal state. Moreover, the pattern of propagation of action across in active systems often resembles a tree structure in which small impulses trigger large reconfigurations. Refering to the entity $a$ under reconfiguration as $a'$, the propagation profile $\zeta(a, \phi)$ at the $k$-th step becomes the hypersurface

\begin{equation}
   \zeta^{D-1}_k(a', \phi) \approx V^\Delta_a \cdot {V^\Delta_a \choose k} \beta^{k}(1 - \beta)^{V^\Delta_a - k}
\end{equation}

with an enveloping hypercircumference

\begin{equation}
   \zeta^{D-2}_k(a', \phi) \approx S^Q_a \cdot {S^Q_a \choose k} \beta^{k}(1 - \beta)^{S^Q_a - k}.
\end{equation}

$\zeta^{D-1}$ and $\zeta^{D-2}$ will be named from hereon as reconfiguration and heat propagators respectively.

We now will proceed to address (c). To do so, we must proceed in two steps. First we must capture the emergence of messengers, which we expect to describe as a the aggregation of the subset of degrees of freedom of an entity into a smaller one with higher energetic stability. Such fragmentation is easily captured in $G$ by having strongly connected components with no edges between them; we will restrict ourselves to interactions where no fragmentation results. First, suppose a fluctuation with sufficiently large magnitude induces a transformation $\bm{T}$ such that

\begin{equation}
    \bm{T}[\phi, G(a)] = (G(a''), G(m)).
\end{equation}

$G(m)$ is the graph of a resulting messenger, a collection of degrees of freedom whose structural complexity should be in proportion to $V^\Delta_a$. Conservation of degrees of freedom implies

\begin{equation}
    E(\phi) + E(a) = E(\Omega_{t + \ell} - \Omega_t) + G(a'') + G(m).
\end{equation}

We may read the latter as follows: a fluctuation acting on $a$ led to the reconfiguration of its internal state into $a''$, dissipating heat and yielding either a new fluctuation $\psi = G(m)$ or a smaller (stable) messenger that, upon contact with other entity, will act as a fluctuation. Not all messengers $m$, however, act equally on all recipients. For instance, molecular binding in proteins is highly specific and depends on conformational changes that act as a sort of signature \cite{savir2007conformational}. Taking this into account, a simple model that generalises the perturbed fraction $f$ for a fluctuation or messenger $\psi$ is the Gompertz function

\begin{equation}
    f(\psi) = C_{\alpha_1, \alpha_2} + \kappa E(\psi) e^{-\alpha_1 e^{-\alpha_2 E(\psi)}}
\end{equation}

where $\alpha_2 = \alpha$ corresponds to reconfigurability the prior model, $\alpha_1$ corresponds to \emph{interaction affinity} and $\kappa$ is the fraction of degrees of freedom in direct contact during an interaction; $C_{\alpha_1, \alpha_2}$ is a constant required such that $f(\emptyset) = 0$.

Observe that we have only described perturbations. We introduce relaxation as the process in which $a'$ regains stability on average. To preserve causality internally, the relaxation process must be somewhat similar than the prior one except that degrees of freedom in $S^Q_a$ are now absent. Hence, it must be the case for its volume propagator

\begin{equation}
   \zeta^{D-1}_k(a'', \phi) \propto \zeta^{D-1}_k(a', \phi)
\end{equation}

Since heat dissipation involves diffusion --i.e. loss of degrees of freedom to $\Omega_t$ in our model, no degrees of freedom are left for $\zeta^{D-2}_k(a'', \phi)$ to operate on. However, the relaxation process cannot occur immediately due to the existence of a difference between the compensation mechanisms that compensate for entropy and from those that compensate for enthalpy \cite{qian1996entropy}. To abstain from the use of time explicitly is to suppose that $\zeta^{D-1}_k(a'', \phi)$ operates always at a distance $d_a$ that depends on the composition of the entity. That is, for all $k$ in $\zeta^{D-1}_k(a', \phi)$, $k' \leq k - d_a$ for $\zeta^{D-2}_{k'}(a'', \phi)$. The quantity $d$ is the minimum relaxation period needed to regain excitability. Thanks to the constant fluctuation of the shape and orientation of $a$, the quantity $d_a$ is intrinsically uncertain up to a error, and mostly so after perturbations have changed the wiring of the internal degrees of freedom. We may characterise empirically this fact by means of the ratio $\xi(a) = d_a/\ell_a$ which we interpret as a dynamical resolution limit of the action of $\psi$ in $a$, and by reasoning, cannot be zero. Hence, the system contains uncertainty at some scale. Based on this, we define $\omega_a = 1/d_a $, i.e. the relaxation frequency. Also, note that the frequency of the entire event is constrained by the longest average relaxation distance, hence

\begin{equation}
    \omega([\cdot]) = \min\left( \langle \frac{1}{d_a} \rangle, \langle \frac{1}{d_b} \rangle \right).
\end{equation}

Let us now introduce $b$ into the picture. To preserve causality, we assume associativity of $\bm{T}$ within a suitable structure $[\cdot]$ which for Fig. \ref{fig:fig_1}B yields

\begin{align}
    \label{eq:perturb-conclus}
    [\phi, G(a), G(b)]  &=[\bm{T}[\phi, G(a)], G(b)] \nonumber \\
                        &= [G(a''), G(m), G(b)] \nonumber \\
                        &= [G(a''), \bm{T}[G(m), G(b)]] \nonumber \\
                        &= [G(a''), G(b''), G(m')].
\end{align}

If $a$ and $b$ are sufficiently complex, causality may established by means the affinity when the process is reversed since, in general, we expect affinities to be reversed. However, there exist two cases where causality may be lost. If the entities are passive, if their propagators are not oriented (i.e. $\beta = 0.5$) and messenger/fluctuation signatures are indistinguishable, then it becomes perfectly legal to relabel $a'' \to a$, $b'' \to b$ and $G(m') \to \phi$ and the analysis is time-symmetric, therefore acausal. In these scenario, either direction or even considering the interaction as two simultaneous events. In the second case, consider a stationary and a moving observer at sufficiently large velocity. It is not hard to image a situation where the direction of motion is such that relaxing and/or propagating portions of all objects (including transitory states $a'$ and $b'$) appear simultaneous to the moving observer, while still being able to assert the irreversible character of the action \cite{israel1979transient}; acausality and hence its interpretation as simultaneity arise again with respect to the order of events. 

Thermodynamically let us consider the effect of $\phi$ on $G(a)$. We expect the internal energy of an entity to be proportional to the number of degrees of freedom in it, thus $U(a) \propto E(a)$. When no perturbation impacts $a$ externally, action still occurs internally to stochastically reconfigure of degrees of freedom without compromising identity; we expect a minimisation of internal entropy with a small fixed generation of heat dissipation $\delta Q^*$ and fixed total entropy generation $\delta \mathcal{S}^*_g$. Also, heat and entropy should grow proportionally with the magnitude of the perturbation. Assume the existence of local entropy generating processes corresponding to an average quantity $\delta \mathcal{S}^{\phi}_g$ operating on each edge in $V^\Delta_a - S^\Delta_q$ during the perturbation-relaxation process and average heat dissipation per edge $\delta Q$. Since 

\begin{equation}
    d \mathcal{S} = \frac{d Q}{T} + d \mathcal{S}_g,
\end{equation}

we can estimate the entropy per internal relaxation step $k$

\begin{equation}
    d \mathcal{S}(a') \approx \mathcal{S}_k(a') \propto \frac{\delta Q \cdot \zeta^{D-2}_k(a', \phi)}{T} + \delta \mathcal{S}^{\phi}_g \cdot \zeta^{D-1}_k(a', \phi).
\end{equation}

and, using a symmetrical argument for $b$ once it is reached by $m$, it must hold for the average entropy produced during the interaction process $[\cdot]$ that

\begin{equation}
    \mathcal{S}([\cdot]) = \int_{t_0(k = 0)}^{t(\ell_a)} d \mathcal{S}_t(a') + \int_{t_0(k = 0)}^{t(\ell_b)} d \mathcal{S}_t(b')+  \approx \sum_{k = 0}^{\ell_a} \mathcal{S}_k(a') + \sum_{k = 0}^{\ell_b} \mathcal{S}_k(b').
\end{equation}

Depending on the variations of phenomena given approximately identical interactions, we expect the effect of $\bm{T}$ to span a distribution of outcomes instead of a single or fixed one. Given the latter, we may more clearly state

\begin{equation}
    \label{eq:int-class}
    \langle [\phi, G(a), G(b)] \rangle = \langle [G(a''), G(b''), G(m')] \rangle
\end{equation}

for stochastically varying $\phi, a$ and $b$. Following the principle of identity of indiscernibles, interactions or rather \emph{interaction classes} must lead to unique descriptions. The uncertainty relations and fluctuations acting upon the configuration of the entities and the classes of entities does not constitute an adequate ground for establishing such identity, since both sides are distributions, and ensembles with similar average observables can exhibit widely different arrays of internal microstates. Defining a class of entity is problematic for the same reason. Many instances of interactions show that $G(m') = \emptyset$ is possible --i.e. all degrees of freedom in $b$ translate either into heat dissipation or $b''$ which, by symmetry, must guarantee the existence of processes where $\phi = \emptyset$. However, the latter does not imply the absence of interactions originating from internal fluctuations in $a$ and $b$; quantum spin measurements are suggestive of this. We thus call interaction sources \emph{extrinsic} whenever $\phi \neq \emptyset$ and \emph{intrinsic} otherwise.

The features that appear to be consistent across all entities discussed so far are their degrees of freedom, frequencies and uncertainties of intrinsic and combinatorial nature. These are, by extension, stochastic quantities and therefore defined by distributions. If the distributions are unimodal and these partition the space of interaction events in a non-degenerate manner, finding a distance metric $\mu$ between interactions is well defined. If  multimodal distributions arise in a given problem, we should expect to find a spectral decomposition of these into simpler --i.e. unimodal- interactions. We immediately recognise the geometry spanned by these distributions as an information geometry \cite{amari2001information}, and the metric $\mu$ as a suitable Fisher metric \cite{costa2015fisher}. Most satisfactorily, doing so translates each interaction as a hypothesis testing instance.

For the principle of identity of indiscernible to hold across interactions, what distinguishes one interaction from another must be the abstract properties of the \emph{mechanism} that operates. Observe that in Eq. \ref{eq:perturb-conclus}, state descriptions of $a'$ and $b'$ have been omitted assuming that they are transient, and therefore somewhat immaterial to the outcomes. This remains so regardless of whether the interaction source is intrinsic or extrinsic. Using these intermediate states brings, however, a source of uncertainty by means of the combinatorial complexity contained in the ensemble of possible future states $a''$ and $b''$ departing from $a'$ and $b'$ correspondingly; note that this view of interactions composes with any intrinsic measurement uncertainty.

Our view is that these along with $m$ contain valuable information about final relaxation states. When $0 \leq U(m) \leq | U(a) - U(a'')| + |U(b) - U(b'')|$, we claim that those that remain in $m$ contain the necessary information to uniquely identify, on average and for a sufficient , both the distribution of its possible origins and the distribution of its possible effects, as well as the properties of the most immediate classes of entities that could have been involved. The argument proceeds on two grounds. First, the statistic complexity of an entity must be in correspondence with the complexity of the ensemble of pattern-generating processes that produced it \cite{emmert2010statistic}. Thus, we are in position to at least infer general properties of the distribution $\langle \bm{T}[\phi, G(a)]\rangle$ and by extension obtain an estimate of $\bm{T}$ via least action. Second, the thermodynamics of requisite variety \cite{boyd2017leveraging} mandates the complexity of $m$ to be commensurate with the amount of memory necessary for pattern-generating processes, and understanding the memory requirements of all messages ``addressed'' to $b$ translate into the general properties of the distribution of patterns $b$ must posses. In this sense, $m$ can be regarded as a specific control system for $b$ that produces a narrow response, and $b$ as the outcome of computing patterns that emerge from sets of messengers that share parts of a common distribution for some $G$. Put bluntly, objects appear to be a consequence of the aggregation and composition of interactions.

The generalized theory of interactions below thus departs from $m \equiv m(a',b')$ as the primary construct in the abstract scaffold corresponding to our attempt to build a parameterizable theory connecting events across scales. Consequently, we state the following guiding principle:

\vspace{10pt}

\textbf{\textit{Principle of Equivalence}}. \textit{Equivalent mechanisms require equivalent messengers. Equivalent messengers describe equivalent interactions}.

\subsection{Interactions}

A \textit{degree of freedom} $\delta$ is a discrete aspect of the state of a system for which the enumeration of a possibly dense range of values --as given by underlying governing laws- yields a description of possible worlds. A possible world is the simplest assignment of one value to the degree of freedom. The range of the degree of freedom is the set containing all values for a given aspect of the world. In our interpretation, the local and thermodynamically irreversible character of interaction events occurs by virtue of interacting systems being holonomic systems subject to multiple forces \cite{giordano2019stochastic}; our description attempts to avoid the complexities of reasoning about dynamics in the resulting Riemannian phase space by focusing on their exchange. Hence, a degrees of freedom corresponds to a generalised distribution.

The \textit{frequency} $\omega$ of an interaction is the minimum of the inverse of the path distance between perturbation and relaxation in the corresponding causal graph $G$ of each interacting system. Consequently, shorter distances produce higher frequencies and longer distances lower frequencies. By convention, frequencies are non-negative quantities, since our description does not involve time-dependent field propagation. Similar to degrees of freedom, frequency describes more properly a portion of the distribution resulting from bounding frequency spectra, describable by means of generalised functions. 

The \textit{uncertainty} $\hbbar$ of a system contains information about the extent of the set of allowable future states of $G$ depending on a particular event and about intrinsic measurement uncertainties present across these macrostates. Since the first component of uncertainty corresponds to the uncertainty relationship between transition speed and thermodynamic cost (of entropic origin) \cite{ito2018stochastic}, and the second to uncertainty as interpreted in quantum phenomena, this quantity is non-negative. To understand where it acquires its character as a distribution, we proceed as follows. Intrinsic uncertainties characterise upper boundaries in acquired knowledge, and correspond to the convex surface that envelopes stochastic measurement outcomes. Extrinsic uncertainty depends on the distribution of states across objects. Since these objects are stochastic themselves, probabilities $p_i$ used in the computation become $\langle p_i \rangle_t$, and naturally 

\begin{equation}
    \mathcal{S} \equiv \langle \mathcal{S}_t \rangle \propto \sum_k \langle p_k \rangle \log \sum_k \langle p_k \rangle,
\end{equation}

hence, $\hbbar$ must be representable using generalised functions.

\begin{figure}[htp]
    \centering
    \includegraphics[width=3.5in]{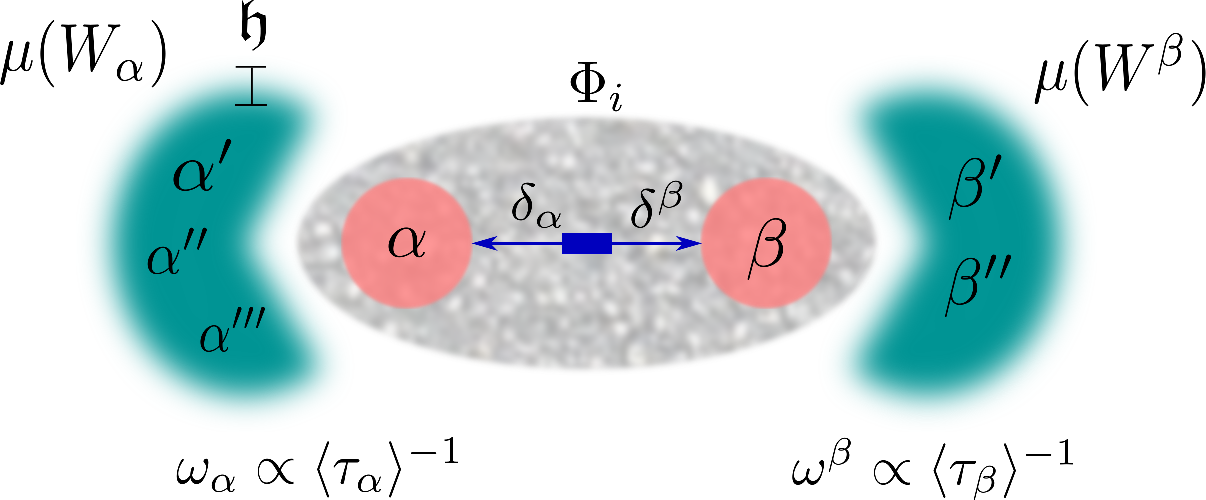}
    \caption{Graphical representation of an interaction in the GToI. Two entities $\alpha$ and $\beta$ embedded in the stochastic context $\Phi_i$ exchange degrees of freedom $\delta_\alpha$ and $\delta^\beta$. In doing so, both undergo relaxation into a set of future macrostates $W_a, W^b$ with frequencies $\omega_a, \omega^\beta$ respectively (proportional to relaxation times $\tau_\alpha, \tau_\beta$). The process is characterised by a generalized measurement uncertainty $\mathfrak{h}$.}
    \label{fig:fig_2}
\end{figure}

An \emph{interaction}, or \emph{interaction class} is a generalised event where action between two systems $\alpha, \beta$ is defined jointly, locally and relationally (Figure \ref{fig:fig_2}). Locally here means that, although not depending on any particular location or \emph{interaction locus}, it can only refer to its relevant neighbouring context $\Phi_i$, where $i$ uniquely identifies the interaction. The event is triggered by either an external (physical or virtual) messenger or a fluctuation, which prompts the exchange of degrees of freedom $\delta_\alpha^\beta$ as described by certain commutation relations with frequency $\omega_\alpha^\beta$ constrained by the structural delay $d_\alpha^\beta$ relaxation of internal structures. Its generalized uncertainty $\hbbar_\alpha^\beta$ is given by an intrinsic uncertainty $\mathfrak{h}$ and a measure $\mu$ over future state diversity. While interactions do not explicitly represent the triggering fluctuation process, it effects are contained in the elements above.

Formally, we start by constructing the manifold from which interactions emerge. First, consider the spaces $\Omega^{\delta}, \Omega^\omega$, and $\Omega^\hbbar$ from which each degree of freedom $\delta$, frequency $\omega$ and uncertainty $\hbbar$ draw their values independent of $\Phi_i$. Here, $\Phi_i$ is the \emph{pseudo-extensive local context}, a structure that represents the media where interactions take place, having its own dynamics, and whose dynamics are in turn impacted by interactions. Consistent with the description so far, $\Phi_i$ is locally labelled by $i$ to signify that only its most relevant extent to the interaction must be considered and that under the principle of background independence it contains a finite and small number of connected degrees of freedom in contact with interacting entities: it provides the convenience of an extensive media but remains local, thus the name pseudo-extensive. The context behaves effectively as a dynamic stochastic network with objects that can have similar properties as a field at the appropriate thermodynamic limit. When $\Phi_i = \emptyset$, interactions are called \emph{fundamental}; conversely, we call a non-empty context a \emph{field network}.

A quantity $\eta_j = \delta, \omega, \hbbar$, in the absence of any other factors, becomes

\begin{equation}
    \eta \equiv X(\Omega^\eta, \Phi_i)
\end{equation}

where $X$ is a random variable whose distribution is $W_{\Phi_i}^\eta$. One can find suitable distributions, amongst which are the thermodynamically irreversible solutions to various Fokker-Planck equations \cite{tome2010entropy,casas2012entropy}. Since we wish to capture the exchange property per degree of freedom, we define the $j$-th degree of freedom as the quantity

\begin{equation}
        \delta_{\alpha,j}^{\beta} = X_{\alpha j}(\Omega^\delta, \Phi_i) + i \cdot X^\beta_j(\Omega^\delta, \Phi_i),
\end{equation}

which can be used to define the \emph{exchanged} degrees of freedom

\begin{equation}
    \bm{\delta}_\alpha^\beta = (\delta_{\alpha,1}^{\beta}, \delta_{\alpha,2}^{\beta}, \cdots, \delta_{\alpha,D}^{\beta})
\end{equation}

for the number of stochastic holonomic degrees of freedom $D$ involved during the interaction. Degrees of freedom donated by $\alpha$ are $\operatorname{Re}(\bm{\delta}_\alpha^\beta)$, and those donated by $\beta$ are $\operatorname{Im}(\bm{\delta}_\alpha^\beta)$, respectively. In certain occasions described below, one may benefit from using reduced degrees of freedom.  Consider $\delta_{\alpha,j} =  \operatorname{Re}(\delta_{\alpha,j}^\beta)$ and $\delta_{j}^\beta =  \operatorname{Im}(\delta_{\alpha,j}^\beta)$ for convenience. Then, these become

\begin{equation}
    \bm{\not \delta}_\alpha^\beta = \sum_j \delta_{\alpha,j} + i \sum_j \delta_{j}^\beta.
\end{equation}

In terms of the dimension associated to the exchanged degrees of freedom, $m$ is an entity where each edge occurs between other entities which are themselves defined via interactions for which other messengers $m'$; that is, the notion of dimension used must be defined in terms of the embeddings $\mathrm{Emb}(\bm{\delta}_\alpha^\beta)$, which contains the next collection of degrees of freedom across all $m'$ implicitly embedded in $m$. At some point, given that the theory is relational, a limit (observable) value $\varepsilon^*$ will be reached for some $\bm{\delta}^* \in \mathrm{Emb}(\bm{\delta}_\alpha^\beta)$. For any degree of freedom $\bm{\delta}$ in $m$, the probability of observing its average value becomes

\begin{equation}
    p(\bm{\delta}) = P(\bm{\delta} = \overline{\bm{\delta}}).
\end{equation}

and the dimension $\bm{d}$ of $\bm{\delta}_\alpha^\beta$ is approximated by the R\'enyi information dimension \cite{renyi1959dimension}

\begin{equation}
    \bm{d}_\gamma(\bm{\delta}_\alpha^\beta) = \lim_{\varepsilon \to \varepsilon^*} \frac{\frac{1}{1 - \gamma} \left( \ln  \sum_{\bm{\delta} \in \mathrm{Emb}(\bm{\delta}_\alpha^\beta)}  p(\bm{\delta})^\gamma \right)}{\ln \frac{1}{\epsilon}}.
\end{equation}

Note that, instead of $\varepsilon \to 0$, the size of the smallest box is finite and given by the magnitude of $\bm{\delta}^*$. A better approximation can be achieved using multifractal analysis \cite{harte2001multifractals}, but the current one suffices conceptually for our purposes. The magnitude of an interaction is a scalar corresponding to the norm $M = || \bm{\delta}_\alpha^\beta ||$, which can be used to obtain a normalised representation $\bm{\tilde{\delta}}_\alpha^\beta = \bm{\delta}_\alpha^\beta/M$ during analysis when only one scale is involved.  Note that for an interaction with symmetric exchange of degrees of freedom and no underlying embedding, we recover the usual notion of dimension as $\bm{d}(\bm{\delta}_\alpha^\beta) = D^2 \bm{\tilde{\delta}}_\alpha^\beta$.

The degrees of freedom of a system must respond to topological constraints during an interaction. While not all degrees of freedom will be constrained, some may. For a subset $J, |J| \leq D$, we express two types of constraints: those pertaining to each side of the exchange, and those pertaining to their mutual transfer. Further suppose without loss of generality that all individual degrees of freedom in $\bm{\delta}_\alpha^\beta$ posses such constraints. Then, if $\bm{\delta}_\alpha^\beta$ can be written as

\begin{equation}
    \bm{\delta}_\alpha^\beta = (y_1 (x_1 \delta_{\alpha,1} + i (1 - x_1) \delta_{1}^{\beta}), y_2 (x_2 \delta_{\alpha,2} + i (1 - x_2) \delta_{1}^{\beta}), \cdots, y_D (x_D \delta_{\alpha,D} + i (1 - x_D) \delta_{D}^{\beta}))
\end{equation}

such that $0 \leq x_j \leq 1$ and $\sum_j y_j = 1$. Under these constraints, the magnitude of the exchange is invariant under transformations $\bm{x}_\nu = \bm{A}^\mu_\nu \bm{x}_\mu$ and $\bm{y}_\sigma = \bm{B}^\rho_\sigma \bm{y}_\rho$. Similarly, fixing constraint yields a partial symmetry over the remaining constraints: we immediately recognise the statement of conservation laws under this new language. The exchange is non-conservative when $|J| = 0$, partially conservative when $|J| \leq D$, and conservative with $|J| = D$. Furthermore, conservation laws obtained in this manner are local, and hence guarantee the existence of at least one causal path involving each conserved degree of freedom as required by Lorentz invariance (special relativity), and permits tracing each path to entities $\alpha$ and $\beta$ consistently having frequencies shifted when exchanged between accelerating inertial reference frames as required by Lorentz covariance (general relativity). We hypothesise that non-conserved exchanges of degrees of freedom arise thanks to incomplete descriptions of $\Phi_i$.

To capture frequencies, we observe that these are also distributions of the form

\begin{equation}
        \bm{\omega}_{\alpha}^{\beta} = X_{\alpha}(\Omega^\omega, \Phi_i) + i \cdot X^\beta(\Omega^\omega, \Phi_i),
\end{equation}

a complex scalar that captures the joint relaxation process after propagation of action across both entities. To draw an analogy to frequency spectra, the value $M$ computed previously acts as an amplitude and the distribution of frequencies for an interaction is analogous to its spectrum. Nevertheless, we observe that frequency density functions are not, in general, continuous.

To obtain a suitable definition for $\hbbar$, we require it to reflect intrinsic and combinatorial sources of uncertainty. For the intrinsic uncertainty, we assume such quantity $\mathfrak{h}_\alpha, \mathfrak{h}^\beta$ can be found per each entity respectively. For the combinatorial part, consider the sets $W_\alpha$, $W^\beta$ that contain all topologically equivalent versions of the entities after relaxation upon interaction associated with outcome distributions. Most generally, we identify the (possibly dense) structure of interactions in the embedding as a random fractal for which the notion of measure has been clarified \cite{taylor1986measure}. Therefore, assuming the existence of a suitable metric $\mu$ such that $\mu(a \cdot b) = \mu(a) + \mu(b)$, one way to compute the uncertainty $\hbbar$ becomes

\begin{equation}
    \bm{\hbbar}_\alpha^\beta = X_{\alpha}(\Omega^\hbbar, \Phi_i)^\mathfrak{h} + i \cdot X^\beta(\Omega^\hbbar, \Phi_i)^\mathfrak{h} = \mathfrak{h} \cdot \left[ \mu \left( W_\alpha  \right) + i \mu \left( W^\beta  \right) \right].
\end{equation}

Furthermore, we can relate $\mathfrak{h}$ to already known quantities. Based on available information, the intrinsic uncertainty must be related to $m$, and more specifically, to dispersion measures of the exchanged degrees of freedom. We choose standard deviation for such purpose. Moreover, since $\operatorname{Re}(\delta_{\alpha,j}^\beta) = \delta_{\alpha,j}$ and $\operatorname{Im}(\delta_{\alpha,j}^\beta) = \delta_{j}^\beta$ are distributions obtained by projection onto components of the complex plane. Let us suppose that the intrinsic uncertainty depends on interdependencies of degrees of freedom contributed by each object, on the interdependencies between each exchange per single degree of freedom, and on interdependencies across different degrees of freedom between objects. This last case corresponds, by construction, to that where conservation laws apply to some degrees of freedom. The form must also be that of an inequality, since intrinsic uncertainty is a lower bound.  Let us propose the ansatz

\begin{equation}
    \label{eq:uncer_1}
    \bm{\mathfrak{h}} \equiv \bm{\mathfrak{h}}_\alpha^\beta \leq \left( \prod_{j, k} \sigma(\delta_{\alpha,j}  \delta_{\alpha,k}) \right) \left( \prod_{j, k} \sigma(\delta_{j}^\beta  \delta_{k}^\beta) \right) \left( \prod_{j, k} \sigma(\delta_{\alpha,j} \delta_{k}^\beta) \right) \left( \prod_j \sigma(\delta_{\alpha,j} \delta_j^{\beta}) \right).
\end{equation}

The first two terms describe the organization of the exchange with respect to $\alpha$ and $\beta$ only, and although we expect them to have an effect on the physics of the interaction, these depend on the inner mechanics driving the interaction. In other situations, the mechanisms may not be transparent for a variety of reasons. Using our Principle of Equivalence, we can at least ascertain that these quantities should be bounded by some common value across equivalent interactions. A similar reasoning can be made for the third term depends on the mechanics at the \emph{locus} of interaction, which includes the context $\Phi_i$. All of the latter are expensive to obtain and require extensive interrogation of interactions by means of \emph{causal experiments}, or experiments where changes in the entities and their corresponding counterfactuals indirectly probe and map mechanisms. The only aspect either directly of efficiently observable through measurements is the exchange of similar degrees of freedom across entities by means of joint experiments. Hence, we may simplify Eq. \ref{eq:uncer_1} as

\begin{equation}
    \label{eq:uncer_2}
    \bm{\mathfrak{h}}  \leq c \cdot \prod_j \sigma(\delta_{\alpha,j} \delta_j^{\beta}),
\end{equation}

where $c$ is an appropriate constant. Consider the situation when conservation laws exist such that a positive, finite local exchange budget $M_{\alpha,j}^\beta$ exists for each $\delta_{\alpha,j}^\beta$ such that $\delta_{\alpha,j} = x_j M_{\alpha,j}^\beta$ and $\delta_j^\beta = (1 - x_j) M_{\alpha,j}^\beta$. This implies that $x_j$ is a distribution, hence Eq. \ref{eq:uncer_2} becomes

\begin{align}
    \bm{\mathfrak{h}}  &\leq c \cdot \prod_j \sigma \left[ \left( M_{\alpha,j}^\beta \right)^2 x_j (1 - x_j) \right] \nonumber \\
                        &= c \cdot \prod_j \left( M_{\alpha,j}^\beta \right)^2 \sigma \left[ x_j - x_j^2) \right],
\end{align}

which for a completely symmetric conservative exchange --i.e. $x_j = 1 - x_j =$ and $M_{\alpha,j}^\beta = M_{\alpha,k}^\beta = M$- becomes

\begin{equation}
    \bm{\mathfrak{h}}  \leq c (\sqrt{2} M^2)^D \cdot \prod_j \sigma^2(x_j).
\end{equation}

In a truly stochastic and irreversible process, the distribution of values in the ensembles spanned by $\delta_{\alpha,j}, \delta_j^{\beta}$ ensures $\sigma > 0$. As we move to a larger scale, the relative variance decreases, and hence the smaller $\sigma$ appears to be. At a sufficiently large limit, $sigma$ becomes negligible. Consider the exchange of position $x$ and momentum $p$ when measuring the state of a wave function $\psi$ using an instrument (Inst), and suppose their decomposition is possible into $\delta_{\psi,x}^{\mathrm{Inst}}$ and $\delta_{\psi,p}^{\mathrm{Inst}}$ respectively. Since, $\psi$ is a distribution, we assume the latter quantities also determine distributions. Using Eq. \ref{eq:uncer_2} 

\begin{equation}
    \bm{\mathfrak{h}}_{\psi}^{\mathrm{Inst}}  \leq c \cdot \sigma(\delta_{\psi,x} \delta_{x}^{\mathrm{Inst}}) \sigma(\delta_{\psi,p} \delta_{p}^{\mathrm{Inst}})
\end{equation}

and observing that

\begin{equation}
    x = c_x \cdot \delta_{\psi,x} \delta_{x}^{\mathrm{Inst}}, \quad p = c_p \cdot \delta_{\psi,p} \delta_{p}^{\mathrm{Inst}}
\end{equation}

must hold for any given exchange where $c_x, c_p$ are determined by the slice in volume of the exchange corresponding to $x, p$ respectively, we obtain the relation

\begin{align}
    \bm{\mathfrak{h}}_{\psi}^{\mathrm{Inst}} &\leq c \cdot \sigma(\delta_{\psi,x} \delta_{x}^{\mathrm{Inst}}) \sigma(\delta_{\psi,p} \delta_{p}^{\mathrm{Inst}}) \nonumber \\
    &= \frac{c}{c_x \cdot c_p} \sigma(x) \sigma(p)
\end{align}

which, after relabelling $\bm{\mathfrak{h}}_{\psi}^{\mathrm{Inst}} = h$ and equating $\frac{c}{c_x \cdot c_p} = 4 \pi$ yields

\begin{equation}
    h \leq 4 \pi \sigma(x) \sigma(p),
\end{equation}

or in its more usual form \cite{ozawa2015heisenberg} with reduced Planck's constant $\hbar = h/2\pi$ 

\begin{equation}
    \sigma(x) \sigma(p) \geq \frac{\hbar}{2}.
\end{equation}

The general argument for classical mechanics was first given by F\"urth \cite{furth1933einige}. Let us now turn our attention to $W_\alpha$ and $W^\beta$. Their internal degrees of freedom of interacting are countable, and their reconfiguration depends on the context $\Phi$ and their governing dynamics. In the interactions view, dynamical laws depend on the composition of interactions within an entity, and so on recursively. To remain tractable, we demand $W$ to quantify reconfigurations of entities in the immediate microscale only. Even so, the task of understanding entities remains momentous from an empirical perspective, since it involves at first value understanding the distribution of coupled responses by means of repeated statistical experiments. Nevertheless, an alternative approach exists: after removing interaction locales within an entity, the state at a sufficiently small time period $\delta \tau$ such that the average structure remains resolvable within intrinsic uncertainty limit $\bm{\mathfrak{h}}$ can be approximated by a collection of basis sets with coefficients corresponding to eigenvalues of the spectral decomposition of its structure. Hence, $W_\alpha$ must be proportional to the number of eigenvalues $\lambda_\alpha$ that span the structure of $\alpha$, which are countable and finite.

Summarising, the existence of an interaction (class) can be appropriately inferred when an exchange of degrees of freedom occurs between two entities mediated by a context, whose structure is in turn explained in terms of interactions, and when frequencies and uncertainties describing the distribution of future entity states can be obtained corresponding to experimental observations. While interaction magnitudes can have any value and/or sign, the number of degrees of freedom involved is taken as positive. Entities are localised interactions that collectively exhibit robustness, and laws are collective manifestations of consequences at some sufficiently large thermodynamic limit. Formally, we denote interactions from here onward as

\begin{equation}
    \intrc{i}{\alpha}{\beta} = (\bm{\delta}_\alpha^\beta, \bm{\omega}_\alpha^\beta, \bm{\hbbar}_\alpha^\beta)_{\Phi_i}.
\end{equation}

\subsection{Interaction spaces}

An \emph{interaction space} $\bm{\mathcal{I}}$ is the fractal \cite{porchon2012fractal}, stochastic \cite{khelif2013stochastic} manifold containing all relevant interactions to describe a given phenomenon at one or more scales. For every $\intrc{i}{\alpha}{\beta} \in \bm{\mathcal{I}}$ exists at least one dense neighbourhood $U$ such that in the structure $(U, \phi: U \to \mathcal{B}^*)$

\begin{equation}
    \mathcal{B}^* = \mathcal{B}^\delta(\mathbb{C}^\mathcal{D}) \times \mathcal{B}^\omega(\mathbb{C}) \times \mathcal{B}^\omega(\mathbb{C})
\end{equation}

where $\mathcal{B}$ denotes a Borel set with a corresponding probability measure and $\mathcal{D}$ is the number of distinct degrees of freedom across all interactions in $\mathcal{I}$. To construct the manifold, we conveniently redefine

\begin{equation}
    \bm{\delta}_\alpha^{\beta,*} = (\delta_{\alpha,1}^{\beta,*}, \delta_{\alpha,2}^{\beta,*}, \cdots, \delta_{\alpha,\mathcal{D}}^{\beta,*})
\end{equation}

and

\begin{equation}
    \intrc{i}{\alpha}{\beta,*} = (\bm{\delta}_\alpha^{\beta,*}, \bm{\omega}_\alpha^\beta, \bm{\hbbar}_\alpha^\beta)_{\Phi_i}.
\end{equation}

where $\delta_{\alpha,j}^{\beta,*} = 0$ in $\intrc{i}{\alpha}{\beta,*}$ whenever  $\delta_{\alpha,j}^{\beta}$ does not appear in $\intrc{i}{\alpha}{\beta}$. Suppose without loss of generality that the probability measure is Gaussian. To every interaction $\intrc{i}{\alpha}{\beta} \in \mathcal{I}$ corresponds a local context $\Phi_i$ belonging to manifold is characterised by neighbourhoods $(U_{\Phi_i}, \varphi: U_{\Phi_i} \to \mathcal{B}_{\Phi_i}^*)$. The state of $\Phi$ resembles that of random media \cite{beran1970mean}, which explains our choice of terminology as a pseudo-extensive field-like entity. We call the association $\intrc{i}{\alpha}{\beta} \mapsto \Phi_i$ an interaction's \textit{local context}, and the association $\intrc{i}{\alpha}{\beta} \mapsto || \bm{\delta}_\alpha^\beta ||$ its magnitude. The projector

\begin{equation}
    \bm{L}(\Phi_i) = \{ \bm{r}_i^\mu \} = R_i(\mathbb{C}^\mu)
\end{equation}

obtains a set of random variables whose distribution is centred on instances of $\intrc{i}{\alpha}{\beta}$.  $\bm{r}^\mu$ corresponds to what we usually interpret as the stochastic position vector, or the \emph{locus of interaction}. The resulting set may be empty, which implies that some interactions are not present in that specific localisation. For example, we only observe interactions between photons and electrons at certain locations and times. Due to the principle of equivalence above, we demand interactions instances within a given distance $\epsilon$ such that $||\intrc{i}{\alpha}{\beta} - \intrc{i}{\alpha}{\beta}|| \leq \epsilon$ to be equivalent, or rather, to choose $\epsilon$ such that the resulting distribution is unimodal so that these belong to the same interaction class. Moreover, we define a \emph{localised interaction} as the pair $(\intrc{i}{\alpha}{\beta}, \bm{r}_i^\mu)$, which describes a \emph{localised interaction space} $\mathcal{I}_r$. When possible, it may be convenient to use interactions with reduced degrees of freedom, or \emph{reduced interactions}

\begin{equation}
    \rintrc{i}{\alpha}{\beta,*} = (\bm{\not \delta}_\alpha^{\beta,*}, \bm{\omega}_\alpha^\beta, \bm{\hbbar}_\alpha^\beta)_{\Phi_i},
\end{equation}

in which case we obtain the \emph{reduced interaction space} $\bm{I}_*$. However, this still proves challenging for visualisation purposes. We can further compute the \emph{scalar interaction space}, a set of real-valued distributions of the form

\begin{equation}
    \sintrc{i}{\alpha}{\beta,*} = || \rintrc{i}{\alpha}{\beta,*} || = (||\bm{\not \delta}_\alpha^{\beta,*}||, ||\bm{\omega}_\alpha^\beta||, ||\bm{\hbbar}_\alpha^\beta||)_{\Phi_i}.
\end{equation}

Observe that $\Phi$ is defined also as product of distributions, and hence must also be a network of more primitive --i.e. fundamental- interactions. We may adopt, for instance, the convention of $\bm{x}^\mu = (x^1, x^2, x^3) = (x, y, z)$. If the distribution is unimodal in $x, y ,z$ then the slicing from which $\Phi_i$ is obtained is spacelike. Within a single interaction space $\mathcal{I}$, localised interaction pairs allow us to obtain

\begin{equation}
    min \left[ \mathcal{L} \left( \Phi_i^{\mathcal{I}}\right) - \mathcal{L} \left( \Phi_j^{\mathcal{I}} \right) \right] = \delta \bm{r}_{ij}^\mu.
\end{equation}

$\delta r = || \delta \bm{r}_{ij} ||$ corresponds to a spacelike distance between two contexts, a distribution describing the minimum resolvable (i.e. unimodal) position change that defines the spatial resolution of the background for an interaction space. A spacelike curve is a path between local contexts $\Phi, \Phi'$ such that every element in the path is also a local context within distance of $\delta r$ one from another. Note that, in general, there exist many possible routes from two different local contexts.

When an interaction space is endowed with a projector $\mathcal{L}$, the resulting inverse fibration is the \textit{localised interaction space} $\mathcal{I}_\mathcal{L}$. Naturally, to each $\mathcal{I}_\mathcal{L}$ corresponds a localised context set $\Phi^\mathcal{I}_\mathcal{L}$ containing at least all local contexts associated with every interaction class in $\mathcal{I}$. Also, $\mathcal{I} = \langle \mathcal{I}_\mathcal{L} \rangle$ and $\Phi^\mathcal{I} = \langle \Phi^\mathcal{I}_\mathcal{L} \rangle$. Consider the mapping between localised context sets $\Phi^\mathcal{I}_\mathcal{L}, \Phi^\mathcal{J}_\mathcal{L}$ and define

\begin{equation}
    \label{eq:eq_time}
    \chi(\Phi^\mathcal{I}_\mathcal{L}) = \Phi^\mathcal{J}_\mathcal{L}  
\end{equation}

whenever every local context in $\Phi^\mathcal{J}_\mathcal{L}$ was produced by a single event leading to an ensemble of relaxation events for each local context instance with frequencies $\omega(\Phi^\mathcal{I}_\mathcal{L}, \Phi^\mathcal{J}_\mathcal{L})$. By setting $\delta \tau = \langle \omega \rangle^{-1}$, we obtain the average proper time required to transform $\mathcal{I}_\mathcal{L}$ into $\mathcal{J}_\mathcal{L}$. The definition of $\delta \tau$ is inherently degenerate, since, it involves multiple ways to construct the localised interaction space and multiple ways to summarise the resulting relaxation frequencies. It is not hard to imagine a summary procedure where the value of the observable depends on the choice of a fixed point $\bm{r}^\mu(\mathcal{I}_\mathcal{L})$, which corresponds to the notion of frame of reference.

\subsection{Aggregation and composition}

\begin{figure}[htp]
    \centering
    \includegraphics[width=2.0in]{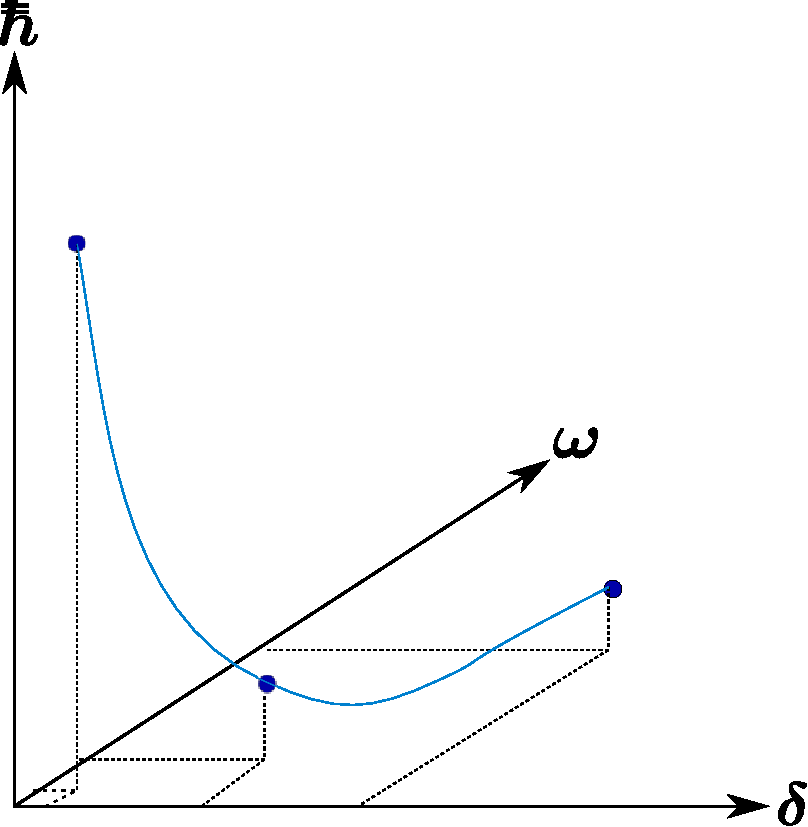}
    \caption{Abstract representation of an interaction space $\mathcal{I}$. This space consists of three interactions (blue points), placed according to their average magnitudes in the space. The interactions are causally connected (blue lines) by means of either aggregation or composition.}
    \label{fig:fig_3}
\end{figure}

Let us now consider various special relations between interactions. Empirically, we observe two general types of associations between them. The first one is aggregation, in which the identity of an entities is preserved under addition of new microscale interactions or subtraction of existing ones within some range of aggregation. For instance, a stream continues to be a stream after removing half its water. The second relation is composition, in which a collection of interactions modulated by aggregation yields new behaviour at a the next macroscale. Superconductivity is an appropriate example of this. While aggregation and composition can be usually expressed in terms of coordinates, the GToI should permit understanding both without requiring localised interaction spaces. In addition, both relations should lead to unveiling causality across interactions in terms of how changes in the interactions involved in aggregation and composition relations translate into changes on those interactions impacted by them. Figure \ref{fig:fig_3} schematises the situation described above.

Aggregation entails an almost total, asymmetric exchange of degrees of freedom where $\operatorname{Re}(\bm{\delta}_\alpha^\beta) = 0$, $V(\bm{\delta}_\alpha^\beta) \approx V(\alpha'')$ and $V(\beta'') = 0$. The identity of one of the interactants is lost ($\beta$), and absorbed by the other one ($\alpha$). By convention, denote the donor entity as $\beta$. As aggregates continue to increase in size, a family of exchanges arises $\bm{\delta}_\alpha^\beta = n \bm{\delta}_\alpha$ with $n$ the number of elementary units $\beta$ involved in the aggregation.  Frequency-wise, each interaction in the family should scale as $\bm{\omega}_{m \alpha}^{n \beta} \propto (\max(m,n))^{-1} \cdot \bm{\omega}_{\alpha}^{\beta}$, since relaxation and propagation are proportional to object length. Observe that the frequency spectra comprises modes spaced at integer intervals.

We find ourselves in a relatively counter-intuitive situation regarding the uncertainty. Since one of the entities of the interaction relaxes into the first one, $W_\alpha$ should be proportional to the volume $V(\alpha'' + \emptyset) \propto V(\alpha) + V(\beta)$, yet the change in the number of eigengraphs describing $\alpha''$ should depend on how the mixing occurs during the merge of degrees of freedom, that is, on the extent of the effective interaction surface involved in the mixing. Since mixing translates into a product due to the new possible combinations between involved surfaces,  

\begin{equation}
    \mu(W_\alpha) \propto \mu([V(\alpha) V(\beta)]^\gamma) = \gamma [ \mu(V(\alpha) +
    \mu(V(\beta))]
\end{equation}

where $0 \leq \gamma \leq 1$ determines the dimensionality of the mixing from being surface-like (i.e. $\gamma \approx 2/3$). Note that the uncertainty of an aggregation can be smaller than that of a regular interaction. This is not expected, since the form of $\bm{\hbbar}$ resembles that of entropy, which is an extensive quantity. But we soon remind ourselves that interactions posses non-extensive properties and, from the point of view of the underlying event, an aggregation tends to involve less information, since it does not change the nature of the entity, only its volume. It is significant to observe that the diameter of $\Phi_i$ is proportional to $n$.

Composition operates radically different. A composition interaction does not necessarily involves the donation of the entire set of degrees of freedom from one object to another. Composition means that new degrees of freedom arise at one scale due to other degrees of freedom being lost at a smaller scale. Similarly, the frequency varies depending on $\alpha$ and $\beta$. However, the key rests upon the uncertainty. Recall that $W_\alpha$ and $W^\beta$ can be expressed in terms of their corresponding bases, related to the subgraphs used to compute the ensemble that defines $\alpha''$ and $\beta''$ respectively. Performing the analysis for $\alpha$

\begin{equation}
    W_\alpha \propto | \langle \lambda_k \rangle_\alpha  |
\end{equation}

such that 

\begin{equation}
    \langle \alpha \rangle = \sum_k \sum_{\delta \in \bm{\delta}_\alpha^\beta} \lambda_k G_{\alpha,k}(\delta).
\end{equation}

An interaction is compositional when, for some set of degrees of freedom involved during the exchange $Y = \{ \delta_{j_1},  \delta_{j_2}, \cdots, \delta_{j_y}\}$ and a new set of degrees of freedom $X$, $|X| \leq |Y|$, the decomposition after relaxation becomes

\begin{equation}
    \langle \alpha \rangle = \sum_k \sum_{\delta \not \in Y} \lambda_k G_{\alpha,k}(\delta) + \sum_j \sum_{\delta \in X} \lambda_j G_{\alpha,k}(\delta),
\end{equation}

which can be interpreted as the loss of existing internal degrees of freedom as determined by the exchange, and the gain of new degrees of freedom in the resulting entities after relaxation. The latter points to the realisation that new degrees of freedom (i.e. edges) arise in the macroscale of $G$ thanks to degrees of freedom being lost at some microscale through compositional interactions. Because of the above, an interaction $\intrc{j}{\mu}{\eta}$ is \textit{more primitive} than an interaction $\intrc{i}{\alpha}{\beta}$ if and only if some of its degrees of freedom appear as parameters in the spectral decomposition of the uncertainty of the latter one, but not in the graph of either resulting entity .

Although we have provided a description of aggregation and composition above, it is insufficient to compute consequences that separate them. For instance, we observe that in general the exchange of degrees of freedom are antisymmetric in one case, and that $W_\alpha$ contains less normal modes, since whole sections of the graph corresponding to various exchanged degrees of freedom have been replaced with fewer, new degrees of freedom. Hence, we are motivated to state our second guiding principle:

\vspace{10pt}

\textbf{\textit{Principle of Composition}}. \textit{For every degree of freedom at once scale a product of compositional interactions must exist giving rise to it}.

\vspace{10pt}

Two corollaries spring from this principle. First, if a degree of freedom $\delta$ is exchanged by an interaction $\intrc{i}{\alpha}{\beta}$ and $\delta$ is explained by interactions $\intrc{k}{\mu}{\nu} \neq \intrc{i}{\alpha}{\beta}$, the interaction $\intrc{k}{\mu}{\nu}$ \textit{is more primitive than} $\intrc{i}{\alpha}{\beta}$, or

\begin{equation}
    \intrc{k}{\mu}{\nu} \rightpitchfork \intrc{i}{\alpha}{\beta}.
\end{equation}

Second, if the interaction exchanges degrees of freedom that arise compositionally from a set containing the same interaction, then it is \textit{fundamental}. This signifies that no further local context exists ($\Phi_i = \emptyset$), and that the concrete theory is background-independent.

The GToI should provide the means to understand at a deeper level the physically possible mappings between interaction spaces, which involves understanding what happens to interactions under replacements of the local context. The following section provides the necessary tools to achieve this.

\subsection{Transformations}

\begin{figure}[htp]
    \centering
    \includegraphics[width=4in]{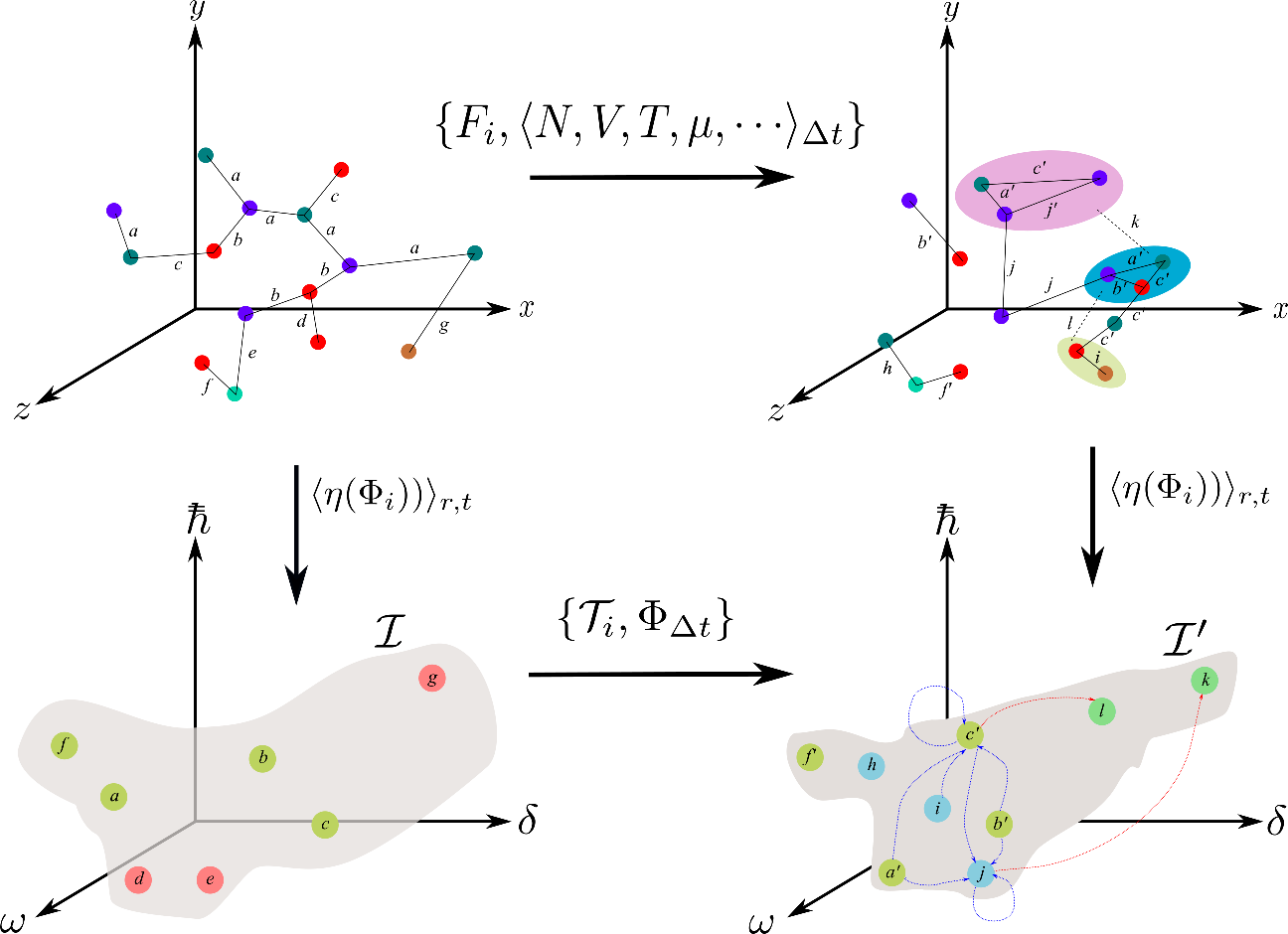}
    \caption{During a transformation, forces applied to a system change microstates and respective macrostate extensive variables. When translated into interactions from a localised interaction space, interaction classes are obtained. Under the effect of a transformation $\mathcal{T}$, tantamount to local corresponding substitution with respect to a degree of freedom by a small amount $\Delta t$, some interactions are preserved and/or translated (yellow green), others disappear (light red) and others arise at the microscale (light blue) and the macroscale (lime green). New microscale (blue line) and macroscale (red line) degrees of freedom arise as a result of composition and aggregation relations. The result of a transformation is to alter the shape of the interaction space, in which boundaries correspond to parametric descriptions taken at various relevant thermodynamic limits.}
    \label{fig:fig_4}
\end{figure}

A transformation interpreted in the context of dynamical manifolds entails a transformation of position, momenta and other degrees of freedom of microscale entities resulting in new values for these, usually constrained by some conservation laws. Concomitantly, extensive macroscale quantities --e.g. $E, S, N, P, V, \mu$- may also change depending on whether the reorganisation of the microscale leads to the emergence of phases with new kinds of properties. In addition, such processes occur during an instant of size $\Delta t$ during which a collection of forces $F_i$ are applied. All the action in the system is bound to some notion of position by means of either a coordinate system or coordinate invariants that enable calculations to be performed. The top half of Figure \ref{fig:fig_4} depicts this notion of transformation.

When looked through the perspective of interactions, we must reason in terms of exchanges of degrees of freedom, frequencies and uncertainties. Let us dissect carefully various situations from this vantage point. First, the application of a force may change the values of spatially dependent degrees of freedom (i.e. $\bm{q}, \bm{p}$). Whether contact forces or fields at at play, a sudden symmetry breaking establishes a preferential direction across some degrees of freedom. If the effect of applying the force (or rather, of the action of the field on the entity) does not change its composition or the types of some interactions involved on average, these interactions are invariant with respect to changes of their local context. Recall that change of the local contexts involves determining the covariance of $(\intrc{i}{\alpha}{\beta}), \Phi_i$. For example, the distribution of interaction classes across loci may vary (reflected in the statistics of $\intrc{i}{}{}$) but the fundamental character of the interaction may not. An interaction can be invariant under substitution of local context.

The second, more interesting case, is that in which interactions are not invariant, as in the bottom half of Fig. \ref{fig:fig_4}. In this case, the effect of a transformation falls into four possibilities.Substituting the local context may result in translating interactions from one location in interaction space to another. Due to the principle of equivalence, the transformation has effectively converted a set of interactions into another.  Another possibility corresponds to a decrease in the number of interactions, since movement across interaction space may unify various interactions into a single one. Finally, a third possibility is the rise of new interactions due to composition. To achieve this, there must exist families of trajectories of local context substitutions with a critical value in which a new type symmetry breaking arises: a single interaction separates into two or more types of interactions along those trajectories. In general, this later aspect reveals another type of irreversibility: when transformation trajectories coincide, information is erased about the past, and when transformation trajectories diverge new information is created. However, since we are in a stochastic interaction space, the inverse transformation taking any of the alternatives back to a point prior to the divergence is guaranteed to differ from it, since the local context will have had undergone variations altering the outcome. Such trajectories describe a type of fractal which can be expected to bear some resemblance around these symmetry breaking points to the resulting geometry of random walks and other stochastic processes on fractal spaces \cite{kanno1998representation,kumagai2004function}.

A transformation $\bm{\mathcal{T}}$ depends on the information stored physically about simultaneous changes in $\intrc{i}{\alpha}{\beta}$ and $\Phi_i$. Such information allows us to refer to the \emph{memory} of the interaction, and the transformation itself as the \emph{robotics} that alters stored information in it. This provides a view of physics that is properly computational, and unveils to some extent why information and energy are intricately connected as suspected since long ago \cite{wheeler1992recent}: because of the construction of the interaction space as a complex stochastic manifold, spatial relations between interactions before and after the transformation are given by the Fisher information metric \cite{calmet2005dynamics} which also corresponds to thermodynamic distance \cite{crooks2007measuring}. At the same time, it extends computation conceptually as a phenomenon rooted in controlling aspects of the fundamental stochasticity present in the mechanical underpinnings of our universe; the type of computation we are used to must be translatable into the GToI as the existence of interactions with familiar properties found across digital systems as exchanges of simple degrees of freedom (usually binary ones) requiring an enormous array of supporting interactions that produce sharp distributions which, at the relevant macroscale of bits, result in deterministic operations. More recently, quantum computation appears to operate by removing layers of interactions to allow the manifestation of stochasticity by allowing superposition and entanglement across Hilbert space and benefit from simultaneous non-determinism. Since the energy of computational interactions should be proportional to the number of constrained degrees of freedom involved during the exchange, more constraints should therefore translate into more expensive computing models; it is thus not surprising why quantum computing elements require lower state-switching energies \cite{gea2003comparison}. We suspect that deeper understanding of the physics of space-time should open the possibility for another type of computation beyond the digital and the quantum mechanical, assuming that quantum gravity is an attainable scientific goal.

To accommodate the richness we suspect exists within interaction spaces without sacrificing clarity, we require mathematical objects capable of nicely hiding the complexity of moving by shifting local context. Essentially, our goal is to specify tensor-like objects, with the condition that they contain functions (or more generally morphisms) instead of having values as entries. To that end, we define a simple homological algebra \cite{cartan1999homological} of tensors of morphisms, or a $\varphi$-\emph{Tensor algebra}.

\begin{definition}[1-morphism]
Let $\varphi: \mathbb{F} \to \mathbb{F}$ be a morphism from a field onto itself. $\varphi$ is a 1-morphism with the following properties:

\begin{enumerate}
    \item The set $\mathfrak{F}$ containing all 1-morphisms endowed with arithmetic operators $(\mathfrak{F}_\varphi, + , \cdot)$  is a field. Operators $-, \div$ can be defined as usual.
    \item $\mathfrak{F}$ contains an \textit{application} operator $\odot$ such that $\varphi \odot a = \varphi(a), \forall a \in \mathbb{F}$. $\odot$ has higher precedence than arithmetic operators.
    \item $\mathfrak{F}$ is a monoid under the \textit{composition} operator $\circ$, having the standard functional interpretation $(\varphi \circ \psi) \odot a = \varphi(\psi(a))$.
    \item For any arithmetic operator $\oplus$, $[\varphi' \circ \varphi \oplus \psi' \circ \psi] \odot a = (\varphi' \circ \varphi) \odot a \oplus (\psi' \circ \psi) \odot a$.
\end{enumerate}

Amongst the members of $\mathfrak{F}$, we find the following special morphism:

\begin{align*}
    0_\varphi \odot a &= 0 (\in \mathbb{F}) \\
    1_\varphi \odot a &= 1 \\
    c_\varphi \odot a &= c \\
    \delta_{ij}^{\varphi} &= 
        \begin{cases}
            1_{\varphi} & i = j \\
            0_\varphi & i \neq j 
        \end{cases}
\end{align*}

\end{definition}

\begin{definition}[$\varphi$-Tensor]
Let $\mathfrak{F}$ be a field of 1-morphisms and $\mu_1, \mu_2, \cdots, \mu_N$ a collection of non-negative indices. a $\varphi$-tensor is a multilinear map of the form

\[
    \bm{T}_{\mu_1 \mu_2 \cdots \mu_N} = (\phi_{\mu_1 \mu_2 \cdots \mu_N}), \quad \phi_{\mu_1 \mu_2 \cdots \mu_N} \in \mathfrak{F}.
\]

Standard tensor index rules apply to $\varphi$-tensors, including contraction rules and tensor arithmentics. The following rules additionally determine $\varphi$-tensors:

\begin{enumerate}
    \item $\bm{u}_\nu = \bm{T}_{\mu \nu} \bm{u}^\mu = \bm{T}_{\mu \nu} \odot \bm{u}^\mu = \sum_i \sum_{j}\bm{e}_i \cdot \varphi_{ij} \odot u_i$, where $\bm{e}_i$ is an element of the basis $B(\mathbb{F}^d)$ of a $d$-dimensional vector space $V(\mathbb{F}^d)$,
    \item $\left( \Sigma_k \bm{T}_{\mu \nu}^{(k)} \right) \bm{u}^\mu = \Sigma_k (\bm{T}_{\mu \nu}^{(k)} \bm{u}^\mu)$,
    \item For $\bm{T}_{\mu \nu} = (\phi_{ij})$ and $\bm{S}^{\nu \sigma} = (\psi_{jk})$, then
        $\bm{U}_\mu^{\sigma,\circ} = \bm{T}_{\mu \nu} \circ \bm{S}^{\nu \sigma} = \sum_j \phi_{ij} \circ \psi_{jk}$.
\end{enumerate}

\end{definition}

Consequently, we define a transformation $\mathcal{T}: \mathcal{I}$ as a discrete operation, where for $\Phi_i^\mathcal{T}$ denotes the state of the local context $\Phi_i$ associated with $\intrc{i}{\alpha}{\beta}$ such that for a sufficiently small difference $\Delta \Phi_i = \Phi_i^\mathcal{T} - \Phi_i$ where $|| \Delta \Phi_i || \geq \bm{\mathfrak{h}}_{\Phi_i}$ the transformation becomes

\begin{equation}
    \bm{\mathcal{T}}(\intrc{i}{\alpha}{\beta}, \Delta \Phi_i) \approx \bm{F} \odot \Delta \Phi_i + \bm{G} \odot \intrc{i}{\alpha}{\beta} + \bm{H} \left[ \bigcirc_{j \in \Lambda} \bm{\Psi}_j \right] \odot \intrc{j}{\mu}{\nu}.
    \label{eq:transformation}
\end{equation}

In Eq. \ref{eq:transformation}, we find that the transformation is approximated by the sum of three $\varphi$-tensors. $\bm{F}$ is the \emph{local context memory $\varphi$-tensor}, containing information about how the components of an interaction vary as a function solely of the difference obtained by replacing one local context with another. Next, $\bm{G}$ is the \emph{internal memory $\varphi$-tensor}, which only depends on the properties of the interaction prior to the local context substitution. The third term contains two $\varphi$-tensors, $\bm{H}$ and $\bm{\Psi}_j$ associated with interactions in a set $\Lambda$. This set contains all interactions that are most immediately primitive with respect to $\intrc{i}{\alpha}{\beta}$,

\begin{equation}
    \Lambda = \{ \intrc{i}{\mu}{\nu} \in \mathcal{I} | \intrc{i}{\mu}{\nu} \rightpitchfork \intrc{i}{\alpha}{\beta} \}.
\end{equation}

The $\varphi$-tensor $\bm{H}$ captures aggregation effects, while $\bm{\Psi}_j$ are $\varphi$-tensors associated to each primitive interaction in $\Lambda$ associated to generative effects \cite{adam2017systems}. For those, some facts become immediately apparent. All  $\bm{\Psi}_j$ must have the same rank $n$ and dimensionality $m$, and their density --fraction of non-zero entries from the $m^n$ possible ones- must be proportional to the complexity of the generative effects. Since the description of transformations across interaction spaces is inherently degenerate, the choice of $\varphi$-tensors, specially those associated with primitive interactions, should be as parsimonious as possible. In many simple cases, we anticipate the transformation to be separable into individual transformations for $\bm{\delta}$, $\omega$ and $\bm{\hbbar}$, in which $\bm{F} \to \bm{f}$ and $\bm{G} \to \bm{g}$ are likely 1-morphism. While it is possible that a single transformation $\bm{\mathcal{T}}$ can correctly capture interaction transformations for an entire interaction space, we expect this to be the case only for very simple systems. In reality, interaction spaces will likely be partitioned into subsets associated with certain transformations.

Up to now, the local context has remained opaque. From Eq. \ref{eq:transformation}, the simplest guess is for it to be a tensor quantity (not at $\varphi$-tensor) that becomes a vector quantity due to index contraction rules. A simple instance of this occurs when

\begin{equation}
    \Phi = \langle \Phi_i \rangle_{\mathcal{I}}
\end{equation}

which we immediately recognise as a \emph{mean-field theory}. These facts justify its \emph{pseudo-extensive} character and, when GToI instances are background-dependent, the field description is appropriate and calculations proceed as described above.  Let us now tackle the case when $\Phi_i$ is itself a network of interactions. Using connectivity information from the network, we perform a first approximation by computing the regular tensor $\bm{X}_{\mu \nu \sigma} = \bm{X}_{\mu \nu \sigma}^{(\bm{\delta})} \otimes \bm{X}_{\mu \nu \sigma}^{(\bm{\omega})} \otimes \bm{X}_{\mu \nu \sigma}^{(\bm{\hbbar})}$ where each component corresponds to the inner product of interaction components per each interaction pair $i,j$. Hence, $\bm{F}$ must be a fourth-rank $\varphi$-tensor. It is interesting to note that, when degrees of freedom in this tensor correspond to spacetime coordinates, the sub-tensor

\begin{equation}
    \bm{\Pi}(\bm{X}_{\mu \nu \sigma}) = \bm{X}_{\mu \nu \sigma}^{(\bm{\delta}(\bm{r}))} \otimes \bm{X}_{\mu \nu \sigma}^{(\bm{\omega})}
\end{equation}

describes a \textit{propagator}, an entity that contains the information about the relation between an interaction event and its surrounding space. For instance, the projector contains attenuation effects due to irreversible loss of degrees of freedom in the interaction to the local context.

Summarising, the theory of interactions defines transformations as linear sets of $\varphi$-tensors containing 1-morphisms that can be composed and applied to interactions. A \emph{concrete theory of interactions} (CToI) is a complete specification of $\mathcal{I}$ by providing its interactions, a description of the local contexts by means of pseudo-field vectors or interaction graphs, the primitiveness relations between them, and the set of transformations that apply to the interactions in terms of $\varphi$-tensors.

\section{Recovering field descriptions}

We are now interested in, given a CToI, recovering the state of a field at a given time and, by extension, at all times of interest. To do so, we first discuss another application of transformations in the GToI. At first, their definition was motivated by asking how an interaction varied under small changes of the local context. Doing so provided us with a recipe to evolve an interaction space based on small steps, which can be defined for single degree of freedom between $\Phi_i$ and $\Phi_i^{\bm{\mathcal{T}}}$. By observing the form of $\bm{\mathcal{T}}$, nothing prevents us from asking what happens to local contexts when we change the interaction associated to it.

In the intuitive notion of field, degrees of freedom are embedded in it, and the magnitude of specific observables should be proportional to the volume occupied by them. Also, the frequencies of the interaction should be transferred to the frequency of relaxation in the field. Using our reasoning above, our interest lies in determining how the field looks like after changing its associated interaction with the restriction that a) the field must preserve its propagator and b) all degrees of freedom of interest in the field must be switched from the interaction to the local context. Let the set $\mathcal{D}$ contain all degrees of freedom of interest, the interaction $\intrc{i, -\mathcal{D}}{\alpha}{\beta}$ the new interaction without the degrees of freedom, we seek for the local context transformation

\begin{equation}
    \bm{\mathcal{U}}(\Phi_i, \intrc{i, -\mathcal{D}}{\alpha}{\beta}) = \Phi_i^{+\mathcal{D}}
\end{equation}

in which $\Delta \Phi_i^{+\mathcal{D}}$ denotes the local context augmented with the relevant degrees of freedom. We must provide a recipe to (approximately) find $\bm{\mathcal{U}}$. If $\Phi_i$ is a tensor, we extend it with the space $\mathfrak{D}$ generated by the degrees of freedom of interest such that $\Phi_i^{+\mathcal{D}} = \Phi_i \otimes \mathfrak{D}$ where the respective tensor components are obtained by inverting $\bm{F}$ in Eq. \ref{eq:transformation} and using the shared components between $\Phi_i$ and $\mathfrak{D}$. If the local context is a network, we obtain the tensor $\bm{X}_{\mu \nu \sigma}$, extend it similarly with $\mathfrak{D}$ using $\bm{F}$ to obtain the extended tensor $\bm{X}_{\mu \nu \sigma}^{+(\mathfrak{D})}$ and back into a graph $\Phi_i^{+\mathcal{D}}$. In the simplest possible way, a new interaction will be found with only the degrees of freedom of interest, and the same relaxation frequency as $\intrc{i}{\alpha}{\beta}$. We observe how our construction is analogous to how modern field theory operates: interactions between fields can be treated as particle-like entities, and interactions manifest as locally correlated field excitations. We note also that, since the field is an ensemble, $\bm{\mathcal{U}}$ may safely be degenerate, since multiple different outcomes converge as a multimodal distribution. Note that the volume added by $\mathcal{D}$ to $\Phi_i$ corresponds to the magnitude $\bm{Gamma}$.

Back to $\bm{\mathcal{T}}$, we observe that one of its predictions is that the identity of an interaction depends on a range of possible $\Phi$'s. Hence, any degree of approximation will depend on how close to the mean of the distribution contained in the interaction we wish to be, as given by fractional powers of the standard deviation $\sigma^q$ with $q$ called here the \emph{quality factor} of the reconstruction. All the latter points once more to how mechanisms characterise interaction classes uniquely, and by extension, the range of contexts by which we may identify them.

Finally equipped with all necessary tools, we describe various types of reconstruction depending on the properties of $\Phi_i$. Since the reconstruction is spatial, we depart from the localised interaction space $\mathcal{I}_\mathcal{L}$. We provide two algorithms for this purpose: one to compute the spatial extent of the interaction space, and another one to compute its temporal aspect. Our choice of reconstruction is na\"ive and inefficient since it requires generating the entire connected lattice of points. On the other hand, its understanding is straightforward. A more efficient and sophisticated method would make use of spectral decomposition and reconstruction methods for graphs \cite{comellas2008spectral}.

We wish to summarise the significance of our model in the context of dynamical systems. First, the GToI preserves the covariance of interactions and their local context. In dynamical systems, one expects field values along with 4-position vectors to fully determine an instant and subsequent events. Since the local context is a distribution whose position projector maps contexts to an information geometry of $\bm{r}^\mu$'s, there exists a correspondence between observed average trajectories and the course of statistical expectation when moving from one interaction space to another in time, while allowing for surprising events to occur. Also, observe that the field descriptions contain propagation and attenuation parameters that correspond to inverse decay laws, and moreover for every decay there exists an information loss mechanism that is thermodynamically irreversible. Finally, we have not lost the relativistic element present in a complete spatio-temporal reconstruction, since a localised interaction must be given to compute the temporal element.

\begin{algorithm}[H]
\SetAlgoLined
\KwData{$\mathcal{I}_\mathcal{L}$, the collection of $\bm{\mathcal{T}}$'s $\bm{\mathcal{U}}$'s associated with all interactions in $\mathcal{I}$, the set of target degrees of freedom $\mathcal{D}$.}
\KwResult{An instantaneous field description for a set of degrees of freedom $\mathcal{D}$.}
 $\bm{I} \leftarrow \emptyset$ \;
 $\bm{dr}^\mu_{*} \leftarrow \min || \bm{L}(\Phi_i) - \bm{L}(\Phi_j) ||$ \;
 Generate a lattice $\mathfrak{L}$ with points $\bm{r}^\eta = \bm{L}(\Phi_i)$ and content $\Phi_i$ spaced according to $\bm{dr}^\mu_{*}$ across $\mu$ dimensions such that all localised interactions are included\;
 Compute the set of all propagators $\mathfrak{P}$ \;
 \ForEach{$\bm{r}^\eta \in \mathfrak{L}$}{
     \eIf{$\bm{r}^\eta == \bm{L}(\Phi_i)$}{
        Replace $\Phi_i$ in $\mathfrak{L}(\bm{r}^\eta)$ by $\bm{\mathcal{U}}(\Phi_i, \intrc{i, -\mathcal{D}}{\alpha}{\beta})$ \;
     }{
        $\mathfrak{L}(\bm{r}^\eta) \leftarrow \Phi_0$, a suitable ground-state local context.
     }
     Endow $\mathfrak{L}(\bm{r}^\eta)$ with $\mathfrak{X}$
 }
 For an arbitrary point in the extent of $\mathfrak{L}$, use propagators in $\mathfrak{P}$ to compute spatially-dependent field decay per degree of freedom.
 \caption{Spatial field reconstruction of a CToI using $\mathcal{I}_\mathcal{L}$.}
\end{algorithm}

\begin{algorithm}[H]
\SetAlgoLined
\KwData{A lattice $\mathfrak{L}$, a next time $\tau$, $\delta \tau$, an interaction $\intrc{i}{\alpha}{\beta}$.}
\KwResult{A new lattice $\mathfrak{L}'$ with changes computed as a f.}
 $s \leftarrow 0$ \;
 \While{$s < \tau$}{
    \ForEach{$\bm{r}^\eta \in \mathfrak{L}$}{
        $\mathfrak{L}'(\bm{r}^\eta) \leftarrow \Pi(\mathfrak{L}(\bm{r}^\eta))$
    }
    $s \leftarrow s + \delta \tau$
 }
 Compute the respective interaction space $\mathcal{J}$ using $\intrc{i}{\alpha}{\beta}$ to set the frame of reference.
 \caption{Temporal field reconstruction of a CToI using $\mathfrak{L}$ from Algorithm 1.}
\end{algorithm}

As a particular ending note for this section, we find the notion of deterministic chaos hard to substantiate under the GToI. Thermodynamic irreversibility mandates stochasticity, which implies that diffusion will make large deviations in spacetime to be unlikely as a function of variance. Hence, the arbitrary dependency of motion on accuracy is entirely removed, since it is not attainable, sustainable or verifiable. In addition, the way the GToI allows field reconstructions is amenable to stochastic amplification \cite{hu1998wave}, a phenomenon that appears to be responsible for observations that resemble chaotic trajectories. Only when scales between phenomena are sufficiently apart --i.e. thermal effects of air on a double pendulum- that trajectories resemble those predicted by deterministic chaos, yet most complex multiscale stochastic systems lack large scale separations. More work is needed on this subject to fully understand the consequences of the GToI on the existing view of the adequacy of chaotic descriptions of phenomena in dynamical systems.

In the following sections, we explore some conceptual consequences of the GToI described above. To do so, we will focus on three central aspects: what the GToI suggests about the character of physical laws --including what they may be and how they arise, how the principle of least action lives in the probabilistic structure of the GToI, and some hints about how a generalized Correspondence Principle may be stated.

\section{GToI in relation to the character of physical laws}

\begin{figure}[htp]
    \centering
    \includegraphics[width=1.5in]{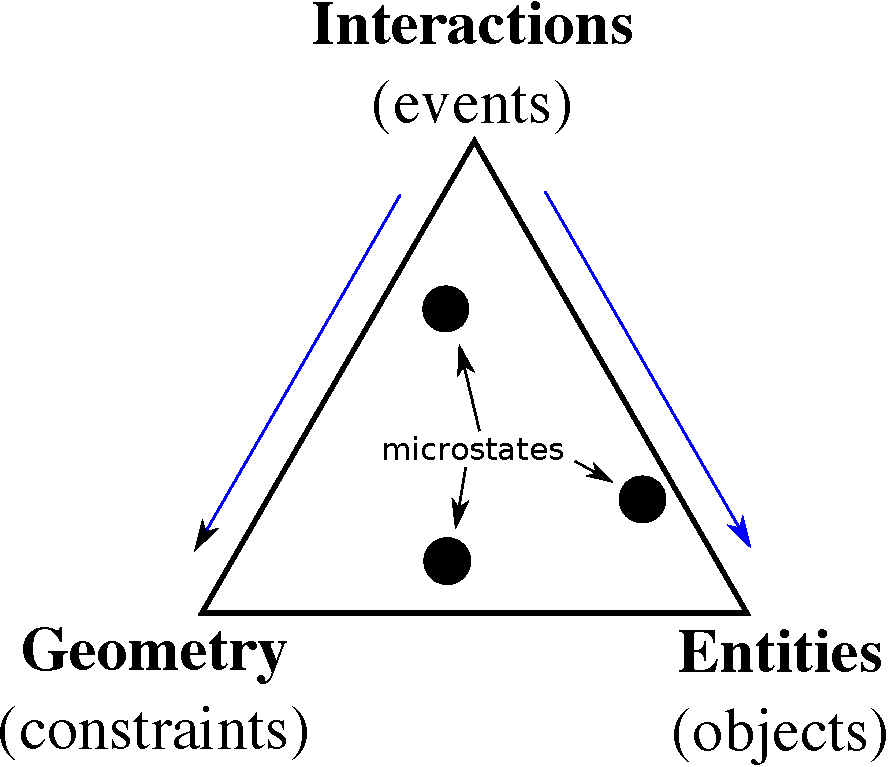}
    \caption{The character of physical laws and objects in the GToI. Interactions are fundamental, and laws (i.e. geometries) and objects are derivative consequences of how interactions aggregate and compose.}
    \label{fig:fig_5}
\end{figure}

The notion of a physical law at the back of our minds is of a universal proposition whose validity holds irrespective of the instance where we look for it. Conservation laws are a magnificent example where all what is required is the presence of a differential symmetry of the action to state the conservation of the respective quantity. A physical law must be sufficiently abstract to describe (almost) all (of) the known universe, and at the same time it must be straightforward enough to remain applicable without much complications. As described in the first article, discovering fundamental rules that apply to the universe across scales appears not to be facilitated by the usual (and convenient) view of dynamical manifolds, in which interactions are only implicitly represented. Figure \ref{fig:fig_5} schematises how objects, laws and microstates are related in the GToI.

The first difference the GToI introduces is to change this situation, and place interactions as entities from which physical laws as we know them can arise naturally. We have shown how the view of interactions are in direct correspondence to mechanisms, and how mechanisms at the most abstract level correspond to hierarchies of interactions. Contrary to other approaches such as master equations, being able to expose mechanisms across scales and allowing them to remain distinguishable can bring a new kind of conceptual simplicity and integration lacking across science domains. On the other hand, gaining knowledge about the structural richness of a complex, thermodynamically irreversible system comes at the expense of loosing information about particular states in favour of distributions, except when averages are available.

The second significant difference is the conceptualisation of how laws operate across multiscale phenomena. Let us consider for a moment the challenges found at the interface between quantum chemistry and molecular dynamics. To enact two different sets of laws, we need to recur in practice to \textit{computationally aphysical} implementations. For instance, one may stop molecular time to recompute self-consistent field theories at the quantum level that are fed back into the specification of the potential energy before proceeding with the integration of the molecular equations of motion later; in many situations, the resulting approximations must make uses of semi-empirical findings. While the GToI would sacrifice the ability to obtain trajectories straightforwardly --still viable yet more involved- its ability to connect events across scales appears to bring the aspiration of an integrated view of physical phenomena a step closer. It is not therefore unreasonable to suppose that theories resulting in aphysical computations may gain significantly to attempts to recast them using the GToI.

A third crucial difference is how the GToI describes the interacting entities. The materiality of objects dominates descriptions across science domains, not just physics. A view of materiality through objects directly involves positions and trajectories, and a linear view of time. Attempting to leave the dynamical view without renouncing some of its principles tends to make matters cumbersome and formally hard. The difficulties, both formal and numerical with the solution of stochastic differential equations, exemplify why introducing further realism without revising assumptions such as the need for an underlying continuous manifold results is major hurdles. In the GToI, \textit{objects are a relational consequence of interactions}, whose materiality only depends on the volume occupied by degrees of freedom in relational space. Note that our definition of volume is state in terms of a general number of dimensions since it corresponds, in reality, to the number of edges of a certain type in a graph.

To further understand the contrast, consider a living cell. Starting at the membrane, interactions with the environment are mediated by interactions between lipids, which organise in bilayers. Within these layers, proteins interact with a host of other molecules. Membranes as a whole interact with concentration gradients at the scale of a single cell, having ion channels and other mechanisms as their most immediate macrostate. Inside the cell, molecules interact to sustain thermodynamic processes that transduce matter and energy intro structure and function. The genetic material stored in DNA with its translation and transcription processes depends on hydrogen-hydrogen interactions, which themselves are rooted in the interactions of orbitals, and those on the interactions of electrons. Of these, we expect electrons to arise from other, more fundamental interactions below currently measurable forms of matter and energy.

In the life of a cell, however, the view of interactions is consistent with thermodynamic irreversibility, with organismal robustness and with the persistence of identity under the constant substitution of matter. Imagine two types of simulations. One would use the usual trajectory and frequency based methods to understand the reality of the cell as various perturbations are performed. The complexity of simulations required to understand the completely localised view of objects in a cell --of \emph{all} relevant objects involved- becomes daunting quickly even without going below a coarse molecular level \cite{bhat2019whole}. Doing so while tracking the association between modules and their parts ends in combinatorial explosions. In the GToI, if we relinquish thinking in terms of objects and their composition and reason in terms of interaction classes, we gain valuable information about interactions, their aggregation and composition, as well as of the distribution of responses a system can produce.

Finally, let us consider the character of physical laws in the GToI. To do so, let us further split the meaning of physical law into two different, yet interlocking components: physical law as \textit{what is allowable to happen in a system}, and physical laws as \textit{the dependency on a large number of entities, events and forces such that recognisably new behaviour appears}. The first one corresponds to the geometry of the system, and the second one to statistical limits.

The first interpretation in the GToI is straightforward, since the geometry of the system is it the information geometry of its interaction space. Since motion across interaction space does not involve time (only differences between local contexts), one can easily imagine computational experiments that map changes across interaction space by sampling the effect of interaction transformations $\bm{\mathcal{T}}$ through a Monte Carlo approach. We envision the procedure to work as follows. A sampling of interactions across interaction space is generated around $\mathcal{I}$ as densely as possible. Using the set of transformations available, $\xi(\mathcal{I})$ is computed until one of two events happens per trajectory: a) the interaction disappears from the space, or b) the interaction splits into two or more interactions. By repeating this process many times and covering the space as much as possible, the geometry of the interaction space can be progressively mapped. We identify the final volume resulting of the evolution of interaction space where the number of trajectories are preserved as embodying the laws of that space. The volume is an open set, yet it is finite for realistic situations. Hence, particular laws arise as a result of fixing an interaction space and mapping its consequences by sampling and evolving it.

The second interpretation, key in the study of complexity, self-organization and emergence, depends on the manifestation of generative effects across systems due to their aggregation and composition. Even the smallest metals showing electrical conductivity are composed by a non-trivial number of atoms \cite{ball2011smallest}, while two or three organisms can easily construct complex routines, coordinate, compete and communicate. Recall from the description above that the volume associated with laws for a given interaction space only covered non-diverging or non-disappearing trajectories. Using the same logic, the envelope of that volume must be related to the manifestation of new laws, since it is what separates them. Furthermore, sheer aggregation of interactions does not appear to have a significant effect for the genesis of new laws; composition, on the other hand has a definitive effect on how a trajectory can be rapidly (and non-linearly) modulated across interaction space. The more interaction classes involved, the richer the behaviour of the system, and the more likely its trajectories to diverge or disappear. Motivated by this, we conjecture the existence of the following law of nature.

\vspace{10pt}

\textbf{\textit{Law of Repertoire Sufficiency}}. \textit{The limit number of microscale entities required for a new law to robustly manifest at its corresponding macroscale is inversely proportional to the number of different interaction classes in the corresponding interaction space, namely}

\[ 
    N_{\mathrm{lim}} \propto |\mathcal{I}|^{-s}
\]

\textit{for some real value $s$.}

\vspace{10pt}

We note that our view includes a tentative answer to the question of whether the laws of nature are immutable of changing \cite{kauffman1997possible, smolin2004cosmological}: not only the laws of nature are far from fixed in the GToI, but they can evolve and be naturally selected. Since fluctuations occur at the base of the hierarchy -- the hypothetical realm of quantum cosmology- it is possible for some of those to explore previously, yet more favourable places within interaction space. Favourable means here stronger invariance under perturbations and transformations. Since these interactions are likely connected upwards to other less primitive interactions in non-trivial manners, these in turn will likely evolve and statistically materialise into other relatively invariant interactions. Unsustainable interactions at any given point either disappear or branch into more stable regimes. As interactions become localised and start permeating usual dynamical manifolds, previous laws are superseded by a sweeping phase transition. We believe this bears some significance for the fine tuning problem in cosmology \cite{smolin2003self}, or the problem of finding mechanisms that explain the precise values of the fundamental constants in our universe.

\subsection{The principle of least action}

The principle of least action takes paramount importance in the history of physics \cite{rojo2018principle}. The action functional contains all the necessary elements to describe and compute consequences for classical, quantum and relativistic systems. It can be extended to dynamical systems with small perturbations by interpreting the geometric consequences of least action in the space of curves as the existence of suitable processes that preserve averages \cite{heymann2008geometric}. From general flows \cite{brenier1989least} to nervous action potentials \cite{dickel1989hamilton}, the principle of least action holds steadfast. Thermodynamically, least action appears to align with the dispersal of energy \cite{annila20102nd}; quantum mechanically, the shortest paths are the most probable ones as given by the solutions of path integrals corresponding to various physical situations \cite{grosche1995solve}.

We can interpret the minimisation of the action as follows using the GToI. Motion across a space or a medium involves interacting across a dynamical path. Every new interaction increases the odds for the object to dissipate heat, losing degrees of freedom to the environment. The longer the path, the more interactions will be encountered, the more heat will be dissipated and the more entropy generated. Put bluntly, the reason we observe instances of the principle of least action according to the GToI is because of the preservation of identity during the selection process that occurs within the stochastic shape ensemble in conjunction with a trajectory full of interactions: after each interaction, either internal degrees of freedom remain somewhat preserved (with the exception of those referring to position and momenta in the case of dynamical manifolds) or they are dissipated. To preserve the identity of an object, the entropy generating processes responsible for re-wiring degrees of freedom internally must approximately preserve it, which must remain approximately so when the hierarchy of interactions that comprises the object interact with their own local context.

Since the shape --i.e. the graph- of an object is stochastic, heat generation will always occur, even of the local context remains the same, while edges are randomly replaced to compensate for lost degrees of freedom. This sort of constitutive equilibrium (a stochastic symmetry of the system) can be broken if the number of degrees of freedom that can be perturbed increases instantaneously and cumulative. The stability of an object under interactions implies the existence of an threshold $\beta^*$ in Eq. \ref{eq:disp_pot} that is characteristic for that object; one way to capture this dependency is considering the interaction space that defines the object $\mathcal{I}_\mathcal{O}$ and find a functional 

\begin{equation}
    \label{eq:beta_pla}
    \bm{R}[\mathcal{I}_\mathcal{O}, \bm{\mathcal{T}}, \intrc{j,k}{\mu}{\nu}] = \beta
\end{equation}

that uses information from that interaction space and its transformations to compute how $\beta$ changes under a sequence of $k$ interactions. One can easily imagine that the length of the sequence of interactions in Eq. \ref{eq:beta_pla} is related to the speed of an object across a world timeline; the existence of $\beta^*$ implies sequences at or below a characteristic length $k^*$. Now, fix the time $\tau_{ab} = \tau_b - \tau_a$ for end and start times between two points $a, b$ and postulate all trajectories travelled across both points. Paths whose length yield values below or equal to $\beta^*$ are identity-preserving; longer paths contain more states in the ensemble that deviate significantly from the original identity due to increases in $\beta$ and are filtered, since they entail different mechanisms. The trajectories we observe contain the least possible action precisely because motion is a selection operator that filters out stochastic shapes on the basis of distance and speed necessary to traverse a dynamical manifold, matching $\beta^*$. Another way to think about it is that nature selects for distances and speeds that preserve $\mathcal{I}_\mathcal{O}$ across repeated interactions; we known them usually as \emph{geodesics}. Hence, the principle of least action is a consequence of stochastic selection ensuring preservation of identity of interaction spaces.

We have, in addition, gained more clarity about the role of entropy as a pervasive guiding force. Entropy provides the variety needed prior to the selection by identity in trajectories. Let us consider two extreme cases of this. One is the case of objects whose microstructure is simple that exhibit high rigidity; their value of $\beta^*$ is high, to the point that it takes a significant number of coordinated interactions to cross the threshold that breaks its identity (e.g. a destructive assay in a solid block of metal). At the other end, consider a soap bubble: it does not take many interactions to break its identity. In the middle, we find self-organising systems and life: they posses the ability to harness incoming interactions and retain their identity by adaptively increasing their interaction repertoire while sustaining low entropy production regimes. We aim to explore both classes of phenomena at depth in subsequent publications. Based on the preceding discussion, we conjecture the existence of a new type of conservation law below.

\vspace{10pt}

\textbf{\textit{Law of Conservation of Identity}}. \textit{For an object $\mathcal{O}$ whose interaction space is $\mathcal{I}_\mathcal{O}$, the set of all possible trajectories between two points $a, b$ within the fixed time interval $\tau_b - \tau_a = \Delta \tau$ is reduced to those where the identity of $\mathcal{I}_\mathcal{O}$ is preserved across repeated interactions. That is, only trajectories proportional to a number of interactions
}

\[ 
    k^* \propto s(a,b) = \int_{\tau_a}^{\tau_b} ds
\]

\textit{are allowable if and only if}

\[
    \bm{R}[\mathcal{I}_\mathcal{O}, \bm{\mathcal{T}}, \intrc{j,k^*}{\mu}{\nu}] \leq \beta^*,
\]

\textit{corresponding to geodesics along $\tau$. Moreover, $\beta^*$ induces an additional symmetry (of identity) under repeated interactions.}

\vspace{10pt}

It is worth noting that thermodynamic irreversibility is essential to obtain the principle of least action within the GToI, and that variation, repetition and selection appear to be universal operators.

\subsection{Generalising the Correspondence Principle}

We wish finally to address the general concern raised by Smolin \cite{smolin1995cosmology} about the preservation of linkages across scales were we to search for the physics of a  universe connected across scales. To achieve explanatory closure, it does not suffice to be able to explain phenomena in the universe using laws that only refer to entities within the universe, but no law should describe a single scale in isolation from the others. The development of the GToI is heavily informed by this particular problem in the light of the difficulties found across multiscale phenomena.

A good example of connection between scales was provided by Bohr \cite{nielsen2013correspondence}, observing that quantum mechanics should reproduce at sufficiently high quantum numbers (e.g. energies), naming it the Correspondence Principle. The connection between expectation values for position and momentum operators to a potential responsible for a force acting on a massive particle was established by Ehrenfest \cite{ehrenfest1927bemerkung}. Quantum decoherence was shown much later to help explain the quantum-classical correspondence in dynamical systems \cite{habib1998decoherence} and more recently found to apply at much more generality \cite{fortin2019correspondence}. The Correspondence Principle can be more generally stated as the existence of bridge equations that, under sufficiently large quantities pertaining to a certain microscale, can be used to obtain average outcomes commensurate to those predicted by laws referring only to entities in the corresponding macroscale; these bridge equations not only depend on the aggregation of many states or entities, but on how they compose in relation to themselves and their context (i.e. environment).

\begin{figure}[htp]
    \centering
    \includegraphics[width=4in]{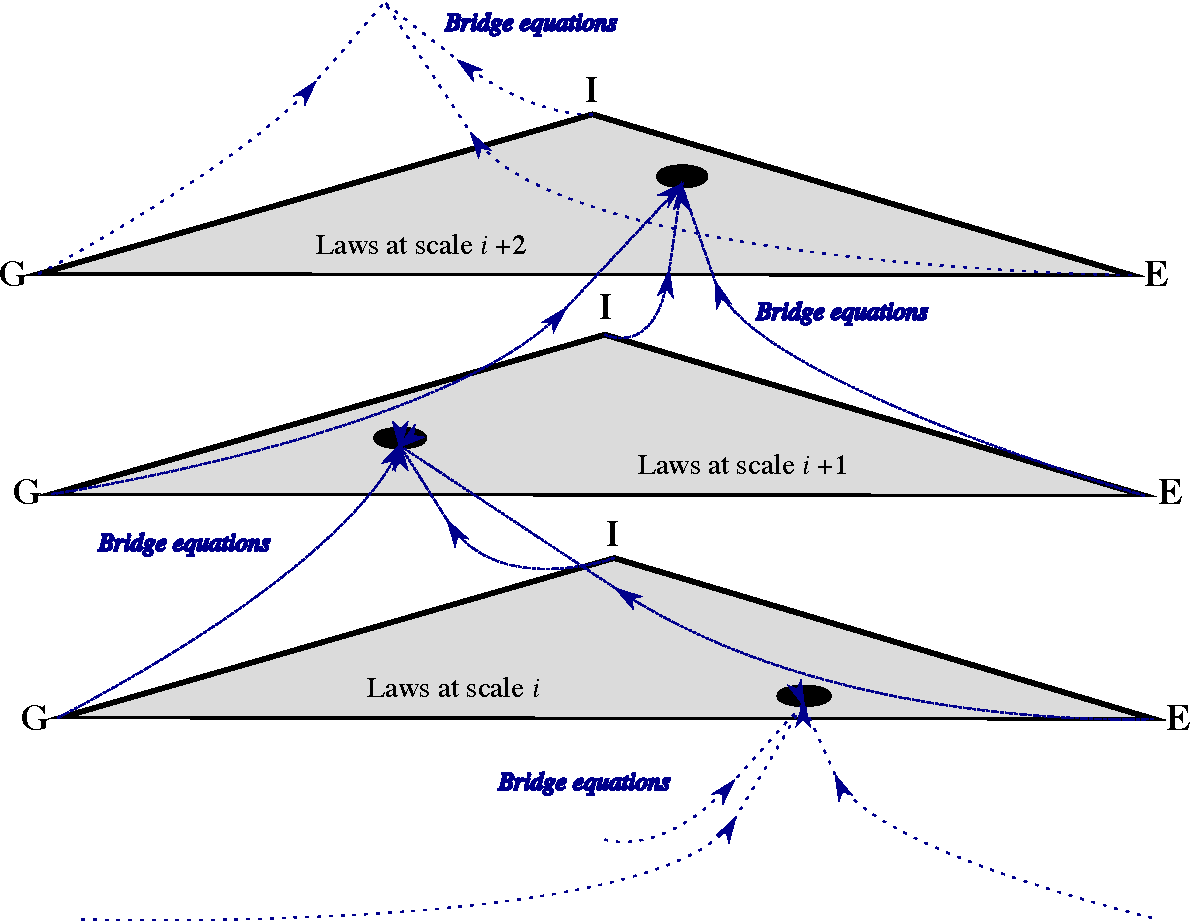}
    \caption{A generalized view of the Correspondence Principle suggested by the GToI. All interactions in the universe are grouped into multiple layers depending on the \emph{more primitive than} order relation. Microstates, laws (i.e. geometries) and objects present up to scale $i$, once their repertoire sufficiency has been reached, produce a new scale with the same formal structure, but different arrangements. Bridge equations arise from the relations of the local context, aggregation and composition present in the transformations pertaining to each interaction space.}
    \label{fig:fig_6}
\end{figure}

Observe that this description is directly encoded into the GToI in various forms. First, the existence of $\varphi$-tensors $\bm{H}$ and $\bm{\Psi}_j$ in Eq. \ref{eq:transformation} captures the general connection between aggregation and composition of interaction classes, shaping the mechanisms expected in the formulation of bridge equations. Second, the effect of the local context is explicit though $\bm{F}$, in which is not hard to imagine situations where properties of $\Phi_i$ induce decoherence. This implies that large variations in the local context erase differences \cite{piechocinska2000information} in the probability distribution governing stochastic processes, leading to average values expected at large limits. Moreover, the evolution of an interaction space towards the appearance and vanishing of interactions as a product of the limiting surfaces as a product of the law of repertoire sufficiency. Since we suppose the universe is connected, we must find a macroscale for every existing microscale. We finally state our extension of the correspondence principle for the general case depicted in Figure \ref{fig:fig_6}.

\vspace{10pt}

\textbf{\textit{Generalised Correspondence Principle}}. \textit{Every microscale $i$ in which limit numbers exist as determined by the Law of Repertiore Sufficiency approximates a corresponding macroscale $i + 1$. Bridge equations between the two are guaranteed to exist solely in terms of relations between local contexts, aggregation and composition of interactions in the interaction space of the microscale.}

\section{Conclusion}

In this article, we have developed the abstract formulation of a Generalised Theory of Interactions. The theory promotes interactions as first-class citizens, and makes use of them as evidence of mechanisms from which governing laws arise. Briefly, an interaction is an exchange of degrees of freedom, mediated by a local environment or context, determined y uncertainty and relaxation relations. Interactions provide the ability to construct a framework that provides explanatory closure, stated only in terms of local events subject to compositionality relations that respect thermodynamic irreversibility at all times while allowing (and to a great extent mandating) causality relations. The GToI, in addition, is modelled upon background independence without excluding cases where field representations constitute a more intellectually efficient pathway. As a useful byproduct, objects and laws become derived entities whose constitution appears to be greatly clarified when looked through the lens of robustness of identity. When applied to a specific CMSS instance, we obtain a concrete theory of interaction (CToI).

Our work, furthermore, contains the statement of various new principles and laws as a means to explore the conceptual possibilities of the theory. First, the equivalence between messengers, mechanisms and interactions as a result of Ashby's law of requisite variety can help, in our current opinion, understand more efficiently CMSS instances at the organismal level and beyond, since probing for mechanisms becomes a tasks of finding the relevant interaction classes and their associated transformations. Second, the compositional construction of degrees of freedom more broadly opens the door to the productive application of powerful mathematical devices that include category theory, topos theory and homotopy type theory; all of these are strongly consistent with the desired computational character of the GToI. Third, the relation between the complexity of a system through the richness of its interaction repertoire appears to bring significant consequences to our understanding of the law of large numbers, and its implications for various types of phenomena including self-organisation and emergence. Fourth, reinterpreting the principle of least action within the GToI results in a plausible mechanism of how mechanics may depend on thermodynamics at a fundamental level. Finally, revisiting the Correspondence Principle provides some hints about the character of requirements needed to achieve explanatory closure in new and existing theories. A curious fact that has not escaped our notice is point particles are not only entirely absent in the GToI, but conceptually incorrect.

Several avenues of investigation remain untouched. A rigorous formalization of the algebra of $\varphi$-tensors must be performed in itself. This is needed in order to understand why, from the universe of such mathematical entities, possibly a subset of them will correspond only to any possible system. While we suspect this is the case in light of the existence of conservation laws, this statement must be rigorously tested, and future work is likely to require the machinery of topos theory or similar to summarise results effectively and efficiently. Another area of work corresponds to the development of the systematics required to find suitable $\varphi$-tensors for specific CToIs; at present, the process is heuristic at best. Furthermore, we envision our preliminary notation to evolve under the pressure of application to useful problems and become more elegant and economic. Rigorous derivations for the principles and laws stated here need to be provided in terms of the underlying algebraic structures. We will dedicate future efforts to address them, and more immediately to apply the GToI to particular cases, including the theory of gases, the foundations of prebiotic evolution and self-organisation.

Finally, the development of the GToI is strongly informed by physics despite the goal, stated at the start, of addressing CMSS at large. Our take is premeditated rather than incidental. By recasting known entities and phenomena for which concepts, principles and laws have been achieved, we have sought to fortify the foundations of the theory to the best of our current possibilities. In doing so, the apparent simplicity of systems and phenomena used in physics unravels into a rich landscape of interactions with new complexities. In a non-trivial manner and to the best of our current knowledge, the universe appears to be a closed, integrated entity in terms of laws and the entities needed to understand it across all scales; in stark contradiction, our theories remain far from such explanatory closure. Understanding such a universe requires placing new demands on how contemporary theories are constructed, their mathematical underpinnings, and their implications for experimental science. Our work suggests that taking seriously such a research program and using complex multiscale stochastic systems as a starting point are conducing to achieving this goal.

\subsection*{Contributions}

S. Núñez-Corrales devised the theoretical formulation. E. Jakobsson edited and revised the manuscript

\subsection*{Competing interest}

Authors declare no competing interest.

\subsection*{Funding}
This work was supported by Illinois Informatics and the ACM/Intel SIGHPC Computational and Data Science Fellowship, 2017 cohort.

\subsection*{Acknowledgments}
This work is dedicated to the memory of late Prof. Eric Jakobsson, who passed away in October 2021.




\bibliographystyle{plain}
\bibliography{references}
\end{document}